\documentclass[onecolumn,11pt,letterpaper]{usetex-onecol}

\usepackage{multirow}
\usepackage{xspace}
\usepackage{enumitem}
\usepackage{algorithm}
\usepackage{algpseudocode}
\usepackage{subfigure}
\usepackage{float}
\usepackage{color}
\usepackage{caption}
\usepackage[T1]{fontenc}
\usepackage{makecell}

\usepackage{times}
\usepackage{breakurl}
\usepackage{color}
\usepackage{enumitem}
\usepackage{graphicx}
\usepackage{xspace}
\usepackage{epstopdf}
\usepackage{url}
\usepackage{cite}
\usepackage[colorlinks=true,linkcolor=black,citecolor=red,pdfborder={0 0 0}]{hyperref}
\usepackage{breakurl}
\usepackage{enumitem}
\usepackage{microtype}
\usepackage{amsmath}
\usepackage{amsthm}

\setlist{nolistsep}
\renewcommand{\paragraph}[1]{\smallskip\noindent{\bf #1}}

\begin{document}

\title{An In-Depth Comparative Analysis of Cloud Block Storage Workloads:
Findings and Implications
\thanks{An earlier version of this article appeared in \cite{li20}.  In this
extended version, we further include the workload traces from Tencent Cloud
Block Storage \cite{zhang20osca} in our analysis.  We extend our findings to
show the commonalities and differences between the cloud block storage
workloads from Alibaba Cloud and Tencent Cloud Block Storage.  }}

\author{Jinhong Li$^\dagger$, Qiuping Wang$^\dagger$, Patrick P. C. Lee$^\dagger$,
	Chao Shi$^\ddagger$\\
$^\dagger$The Chinese University of Hong Kong\\
$^\ddagger$Alibaba Group}

\maketitle

\begin{abstract}
Cloud block storage systems support diverse types of applications in modern
cloud services.  Characterizing their I/O activities is critical for guiding
better system designs and optimizations.  In this paper, we present an
in-depth comparative analysis of production cloud block storage workloads
through the block-level I/O traces of billions of I/O requests collected from
two production systems, Alibaba Cloud and Tencent Cloud Block Storage. We
study their characteristics of load intensities, spatial patterns, and temporal
patterns.  We also compare the cloud block storage workloads with the notable
public block-level I/O workloads from the enterprise data centers at Microsoft
Research Cambridge, and identify the commonalities and differences of the
three sources of traces.  To this end, we provide 6 findings through the
high-level analysis and 16 findings through the detailed analysis on load
intensity, spatial patterns, and temporal patterns.  We discuss the
implications of our findings on load balancing, cache efficiency, and storage
cluster management in cloud block storage systems. 
\end{abstract}

\section{Introduction}
\label{sec:intro}

Traditional desktop and server applications, such as virtual desktops,
operating systems, web services, relational databases, and key-value stores, 
are now moving to the cloud.  {\em Cloud block storage} systems \cite{meyer08,
mickens14, aws_ebs, alicloud, li19, zhang20pbs, zhang20osca} form an
infrastructure that allows cloud service providers to manage large-scale
physical storage clusters.  They also provide virtual disks, referred to as 
{\em volumes}, for clients to host various types of applications.  To allow
performance optimizations and efficient resource provisioning of cloud block
storage systems, it is critical to characterize and understand the I/O
behaviors of the applications in production environments. 

Several field studies have analyzed the I/O behaviors of various architectures
via the collection and characterization of block-level I/O traces
\cite{ahmad07, kavalanekar08, narayanan08, zhou15, harter16, lee17}.  In
particular, the public block-level I/O traces released by Microsoft Research
Cambridge \cite{narayanan08} have received wide attention from researchers and
practitioners.  The traces, which we refer to as {\em MSRC}, have been
extensively analyzed to motivate storage system designs and optimizations,
such as I/O scheduling \cite{narayanan08, liu12, cai15, li16}, caching
\cite{soundararajan10, saxena12}, erasure-coded storage \cite{chan14,
zhang20pbs}, as well as cloud block storage \cite{li19}.  

However, the MSRC traces, which were collected from enterprise data centers
more than a decade ago, may not necessarily reflect the actual I/O behaviors
of today's cloud block storage systems.  Modern cloud environments often host
much more diverse types of applications, some of which feature unique
characteristics (e.g., short-lived tasks \cite{mishra10}) that are not
commonly found in traditional data center environments.  Also, the workloads
in MSRC are generally read-dominant \cite{narayanan08}, while the workloads in
cloud environments are often write-dominant due to the heavy use of read
caches in cloud applications \cite{liu19,wang20}.  Such mismatches motivate
the need of collecting and analyzing comprehensive block-level I/O traces from
real-world cloud block storage systems in large-scale production. 
  
In this paper, we present an in-depth comparative study on the block-level I/O
traces from two production cloud block storage systems. The first set of
traces, which we refer to as {\em AliCloud}, is collected by ourselves from a
production cloud block storage system deployed at Alibaba Cloud \cite{li20}
and covers one-month I/O activities of 1,000 volumes.  The second set of
traces, which we refer to as {\em TencentCloud}, is collected from the Tencent
Cloud Block Storage by Zhang et al. \cite{zhang20osca} and covers I/O
activities of 4,995 volumes over around nine days. 
Both sets of traces feature a large data scale,
totaling billions of I/O requests and hundreds of terabytes of I/O traffic.
We compare the cloud block storage workload characteristics of both AliCloud
and TencentCloud traces with those of the MSRC traces, and identify the
commonalities and differences of the three sources of traces.  To this end, we
provide 6 findings through the high-level analysis on the basic
I/O characteristics of the traces, and further provide 16 findings
through the detailed analysis on the I/O behaviors in terms of the load
intensities, spatial patterns, and temporal patterns.  We provide insights into
load balancing, cache efficiency, and storage cluster management in cloud
block storage systems.  Note that Zhang et al. \cite{zhang20osca} mainly use
the TencentCloud traces for designing efficient cache allocation schemes in
cloud block storage systems, but do not provide an in-depth analysis on the
TencentCloud traces.  To the best of our knowledge, compared with prior
measurement studies on block-level I/O traces \cite{ahmad07, kavalanekar08,
narayanan08, zhou15, harter16, lee17, yadgar21} (see details in
\S\ref{sec:related}), our trace analysis is one of the largest measurement
studies on block-level I/O traces reported in the literature.  We make the
source code of all our analysis scripts available at:
\href{http://adslab.cse.cuhk.edu.hk/software/blockanalysis}%
{\bf http://adslab.cse.cuhk.edu.hk/software/blockanalysis}.

We highlight some major findings of our trace analysis. From the high-level
analysis, small I/O requests dominate in all traces. Both AliCloud and
TencentCloud are write-dominant, while MSRC is read-dominant. For load
intensities, all traces show similar amounts of I/O traffic, while AliCloud
and TencentCloud show more diverse burstiness across volumes and have higher
activeness than MSRC.  For spatial patterns, all traces
show aggregations of reads and writes in small working sets. 
In particular, TencentCloud shows the highest level of
aggregations of reads, implying a more skewed access pattern in reads.  All
traces also show high fractions of random I/Os and varying patterns in the 
update coverage across volumes.  For temporal patterns, all traces have
varying temporal update patterns across volumes and different access
tendencies for the written blocks. For example, each written block in AliCloud
and TencentCloud is likely to be followed by a write, while that in MSRC is
about equally likely to be followed by either a read or a write.  

The rest of the paper proceeds as follows. 
In \S\ref{sec:background}, we present our cloud block storage architecture and
its design considerations.  
In \S\ref{sec:trace}, we introduce the traces for our analysis, and present
6 findings via our high-level analysis. 
In \S\ref{sec:finding}, we conduct an in-depth analysis and provide 16
findings on load intensities, spatial patterns, and temporal
patterns. We emphasize the commonalities and differences of the findings
between AliCloud and TencentCloud. We further discuss the implications of our
findings in terms of load balancing, cache efficiency, and storage cluster
management in cloud block storage systems.  In \S\ref{sec:related}, we review
related work.  In \S\ref{sec:conclu}, we conclude the paper. 

\section{Background and Methodology}
\label{sec:background}

We introduce the cloud block storage architecture considered in the paper
(\S\ref{subsec:arch}).  We further elaborate on how our trace
analysis should characterize the I/O activities in response to the design
considerations for cloud block storage (\S\ref{subsec:motivation}).

\subsection{Cloud Block Storage}
\label{subsec:arch}

\begin{figure}[!t] 
\centering
\includegraphics[width=4in]{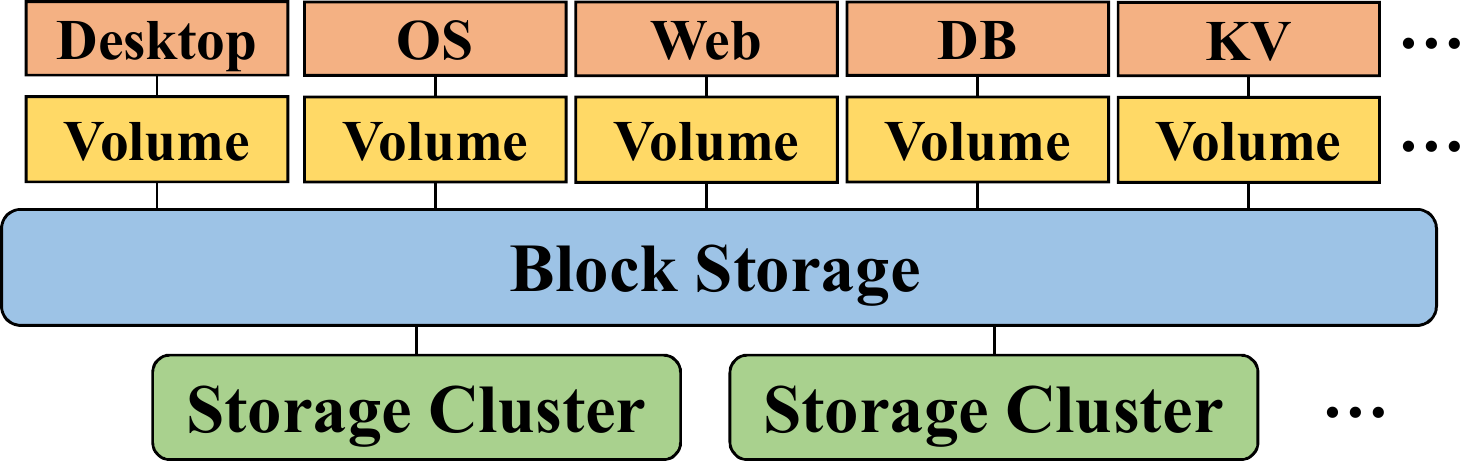}
\caption{Architecture of a cloud block storage system. It
comprises multiple volumes that host a mix of cloud applications (e.g.,
virtual desktops, operating systems, web services, relational databases, and
key-value stores).}
\label{fig:architecture}
\end{figure}

Figure~\ref{fig:architecture} depicts the architecture of a cloud block
storage system considered in the paper.  The cloud block storage
system serves as a middleware layer that bridges: (i) the virtual
disks (referred to as {\em volumes}) that are perceived by
upper-layer applications, and (ii) the storage clusters that provide physical
storage space owned by cloud service providers.  Each application is allocated
with a dedicated volume.  It issues read or write requests through the
dedicated volume to the storage clusters.  Each volume is typically replicated
across multiple storage clusters for fault tolerance.  For performance and
reliability, today's storage cluster are often backed by flash-based
solid-state drives (SSDs) instead of hard disk drives (HDDs) \cite{xu19,
maneas20, yadgar21, han21}. 

In production, a cloud block storage system may manage diverse types of
upper-layer cloud applications (Figure~\ref{fig:architecture}).  The I/O
characteristics of such applications are often largely different, as we show
in this paper. 

\subsection{Analysis Methodology}
\label{subsec:motivation}

Cloud block storage systems should maintain quality-of-services guarantees
(e.g., low-latency requests and fairness) and efficient resource utilizations
(e.g., long device lifetime).  We highlight three design considerations for
cloud block storage systems, namely {\em load balancing}, {\em cache
efficiency}, and {\em storage cluster management}.   In the following, we
explain how each of the design considerations can be addressed in our trace
analysis of I/O activities in cloud block storage. 

\paragraph{Load balancing.}  Maintaining load balancing across storage devices
is important for availability and performance.  If load imbalance
exists, some storage devices may be overloaded by a large number of I/O
requests and cannot serve incoming requests in a timely manner, thereby
increasing the overall I/O latencies.  In addition, the overloading of I/O
requests may aggravate flash wearing \cite{xu19}, leading to reduced endurance.
Since load balancing addresses the performance differences due to the uneven
distribution of I/O traffic, our trace analysis should examine the load
intensities of I/O traffic.

\paragraph{Cache efficiency.} To speed up I/O performance, storage systems
typically cache frequently accessed data based on efficient resource
allocation schemes and admission policies \cite{arteaga16,wang18}.  However,
the high variations of I/O characteristics may introduce improper cache space
allocation and cache management policies, which degrade hit ratios and
increase the overall I/O latencies.  To investigate how the caching design can
leverage workload characteristics, our trace analysis should address the
spatial and temporal aggregations of I/O traffic. 

\paragraph{Storage cluster management.} Enterprise storage clusters
increasingly move to flash-based storage, which is sensitive to varying
workload patterns in both performance and endurance.  In particular, the
update patterns can determine the effectiveness of garbage collection and
wear-leveling in flash \cite{he17}.  Storage cluster management should address
the variations of workload patterns, so as to maintain high performance and
endurance of the underlying flash devices.  Thus, our trace analysis
should focus on the spatial and temporal patterns for update requests.  Also,
as small and random I/Os can degrade the performance and endurance of flash
storage \cite{min12}, our trace analysis should also examine the randomness of
I/Os. 

\section{Traces}
\label{sec:trace} 

We describe the three sets of traces used in our analysis and state the
limitations of our trace analysis (\S\ref{subsec:overview}). We then present a
high-level analysis on the basic statistics as well as the commonalities and
differences on all three traces (\S\ref{subsec:highlevel}). 

\subsection{Trace Overview}
\label{subsec:overview}

Our trace analysis is based on three sets of block-level I/O traces collected
from different production environments. For brevity, we refer to the traces as
{\em AliCloud}, {\em TencentCloud}, and {\em MSRC} in short in the following
discussion.

\paragraph{AliCloud.} The traces were collected by ourselves from a cloud
block storage system deployed at Alibaba Cloud over a one-month period in
January 2020.  The traces are now released at \cite{alicloudtraces}.  They
comprise block-level I/O requests collected from 1,000 volumes, each of which
has a raw capacity from 40\,GiB to 5,000\,GiB.  The workloads span diverse
types of cloud applications (\S\ref{subsec:arch}). Each collected I/O request
specifies the volume number, request type, request offset, request size, and
timestamp (in units of microseconds).

\paragraph{TencentCloud.} The traces were collected from the cloud block
storage system at Tencent Cloud Block Storage \cite{zhang20osca} from
12:00~AM on October 1, 2018 to 1:00~AM on October 10, 2018 in the GMT+8 time
zone (i.e., 9.04~days in total); note that the requests are missing between
1:00~AM and 2:00~AM on October 8, 2018.  The traces can be downloaded from the
SNIA IOTTA repository \cite{tencentcloudtraces}.  They comprise block-level
I/O requests collected from 4,995 volumes.  The workloads are based on
a mixture of cloud applications, including applications dominated
by random accesses and applications with large amount of I/O activity
\cite{zhang20osca}. Each collected I/O request contains the volume number,
request type, request offset, request size, and timestamp as in the AliCloud
traces, except that the timestamp in the TencentCloud traces is in units of
seconds.  However, the TencentCloud traces do not provide the raw capacities of
individual volumes. 

\paragraph{MSRC \cite{narayanan08}.}  The traces were collected by Microsoft
Research Cambridge from a data center of Microsoft Windows servers over a
one-week period in February 2007 and can be downloaded from the SNIA IOTTA
repository \cite{msrctraces}.  They comprise block-level I/O requests
from 36 volumes over 179 disks in 13 servers.  The workloads span a variety of
applications, including home directories, project directories, web services,
source control, media services, etc.  Each collected I/O request includes the
volume number, request type, request offset, request size, and timestamp as in
the AliCloud and TencentCloud traces; it also includes the response time of
the request.  Both the timestamp and the response time are specified in units
of 100\,ns, based on the Windows Filetime timestamp format used by Microsoft
Windows servers. 

\paragraph{Limitations of our trace analysis.} Our trace analysis has several
limitations.  First, both the AliCloud and TencentCloud traces do not record
the response times of the I/O requests as in MSRC, so we cannot conduct
latency analysis on I/O requests in actual deployment.  In particular, the
timestamp field in the TencentCloud traces is in units of seconds, so it is
difficult to conduct fine-grained analysis on the inter-arrival times of
requests in the TencentCloud traces.  Also, both traces do not
indicate the specific applications running atop individual volumes, so we
cannot investigate the relationship between specific application workloads and
their I/O patterns.  Furthermore, all three traces do not include 
the caching policies, cache hit/miss ratios (i.e., we cannot study the impact
of caching in actual deployment) and the age or usage of each volume (i.e., we
cannot study the relationships between I/O patterns and volume ages).
Finally, all three traces do not capture the information of physical storage
devices (e.g., data placement and failure statistics), so we cannot study the
performance and reliability correlations in physical storage. 

\subsection{High-level Comparative Analysis}
\label{subsec:highlevel}

We now present a high-level comparative analysis on AliCloud, TencentCloud,
and MSRC by collectively analyzing the I/O requests of all volumes 
in each set of traces and presenting the overall basic statistics.  Our goal
is to examine the basic properties of the cloud block storage workloads in
AliCloud and TencentCloud as well as the classical enterprise data center
workloads in MSRC.  To this end, we identify the commonalities and differences
of all three traces. 

Table~\ref{tab:basic} summarizes different categories of basic statistics in
AliCloud, TencentCloud, and MSRC, including: (i) the numbers of reads and
writes, (ii) the total amounts of data read, written, and updated, as well as
(iii) the {\em working set sizes (WSSs)} of reads, writes, and updates (an
{\em update} request refers to a write request to a block that has been
written at least once).  We define the total read, write, and update WSSs as
the numbers of unique logical addresses being accessed (i.e.,
read, written, and updated, respectively) by all I/O requests of the traces
multiplied by the block size 4\,KiB. Table~\ref{tab:property} 
further summarizes the key properties of all three traces observed in our
high-level analysis. 

\begin{table}[!t]
\small
\centering
\renewcommand{\arraystretch}{1.15}
\begin{tabular}{c|c|c|c}
\hline
 & {\bf AliCloud} & {\bf TencentCloud} & {\bf MSRC} \\ 
\hline
\hline
{\bf \#Volumes} & 1,000 & 4,995 & 36             \\ 
\hline
{\bf Duration (days)} & 31 & 9.04 & 7           \\ 
\hline
{\bf \#Reads (millions)} & 5,058.6 & 10,030.2 & 304.9     \\ 
\hline
{\bf \#Writes (millions)} & 15,174.4 & 23,592.0 & 128.9   \\ 
\hline
{\bf Read Traffic (TiB)}  & 161.6 & 282.3 & 9.04      \\ 
\hline
{\bf Write Traffic (TiB)}  & 455.5 & 837.2 & 2.39     \\ 
\hline
{\bf Update Traffic (TiB)}  & 429.2 & 804.2 & 2.01    \\ 
\hline
{\bf Total WSS (TiB)}  & 29.5 & 38.7  & 2.87   \\ 
\hline
{\bf Read WSS (TiB)}  & 10.1 & 14.6   & 2.82   \\ 
\hline
{\bf Write WSS (TiB)}  & 26.3 & 33.0  & 0.38   \\ 
\hline
{\bf Update WSS (TiB)}  & 18.6 & 21.2 & 0.17   \\ 
\hline
\end{tabular}
\vspace{3pt}
\caption{Basic statistics of AliCloud, TencentCloud, and MSRC.}
\label{tab:basic}
\vspace{-6pt}
\end{table}

\begin{table}[!t]
\small
\centering
\renewcommand{\arraystretch}{1.15}
\begin{tabular}{c|c|c|c}
\hline
 & {\bf AliCloud} & {\bf TencentCloud} & {\bf MSRC} \\ 
\hline
\hline
{\bf Read WSS over total WSS (A.1) } 
    & 34.3\%  & 37.6\%  &  98.4\% \\ 
\hline
{\bf 75th percentiles of read/write sizes (A.2)} 
    & 12\,KiB/16\,KiB  & 32\,KiB/12\,KiB  & 64\,KiB/20\,KiB           \\ 
\hline
{\bf Volumes with short active periods (A.3) } 
    & 15.7\% & 1.7\%  & 0.0\%         \\ 
\hline
{\bf Write-dominant volumes (A.4)} 
    & 91.5\% & 92.3\% & 52.8\%        \\ 
\hline
{\bf Higher fractions of WSS in larger volumes (A.5)} 
    & Yes & - & - \\ 
\hline
{\bf Larger requests in larger volumes (A.6)} 
    & Yes & - & - \\ 
\hline
\end{tabular}
\vspace{3pt}
\caption{Summary of the key properties of AliCloud, TencentCloud, and MSRC in
Findings~A.1-A.6.}
\label{tab:property}
\vspace{-6pt}
\end{table}

In terms of scale, both AliCloud and TencentCloud have much larger
scale than MSRC in various aspects, including the number of volumes, the
trace duration, the total number of I/O requests, and the total I/O traffic
size.  For example, AliCloud contains 20.2~billion I/O requests, 46.6$\times$
the total number of I/O requests in MSRC.  It also has more volumes
(27.8$\times$), a larger total I/O traffic size (54.1$\times$), and a larger
WSS (10.3$\times$), compared with those in MSRC.  TencentCloud has an even
larger scale than AliCloud in terms of the number of volumes (5.0$\times$),
the total number of I/O requests (1.7$\times$), and the total I/O traffic size
(1.8$\times$), except that its duration only lasts for 9.04 days.  In the
following, we elaborate the I/O characteristics of the three traces. 

\paragraph{Finding~A.1:} {\em Reads span a small proportion of working sets in
both AliCloud and TencentCloud.} 

Referring to Table~\ref{tab:basic}, reads in AliCloud and TencentCloud only
occupy 34.3\% and 37.6\% of the total WSS, respectively, while reads in MSRC
occupy a much larger proportion (98.4\%) of the total WSS.  On the other hand,
writes in AliCloud and TencentCloud occupy 89.4\% and 85.2\% of the total WSS,
respectively. The results indicate that a substantial amount of written data
is never read again in both AliCloud and TencentCloud.  One possible reason is
that some applications tend to only write data but rarely read data (e.g.,
backups or journaling), although we cannot identify the specific applications
running on individual volumes (\S\ref{subsec:overview}). 

\paragraph{Finding~A.2:} {\em Small-size I/Os dominate in all AliCloud,
TencentCloud, and MSRC.}

\begin{figure}[!t] 
\centering
\begin{tabular}{@{\ }c@{\hspace{0.2in}}c}
\multicolumn{2}{c}{\includegraphics[width=3.3in]{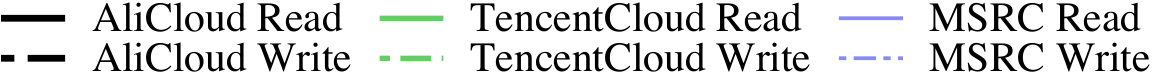}}\\
\includegraphics[width=2.25in]{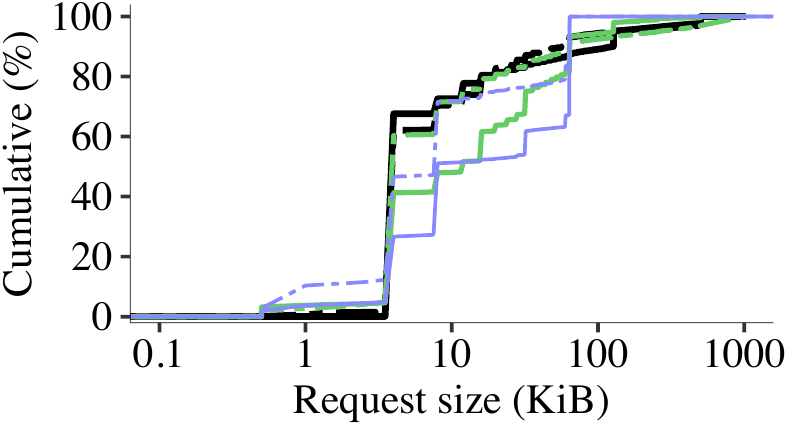} &
\includegraphics[width=2.25in]{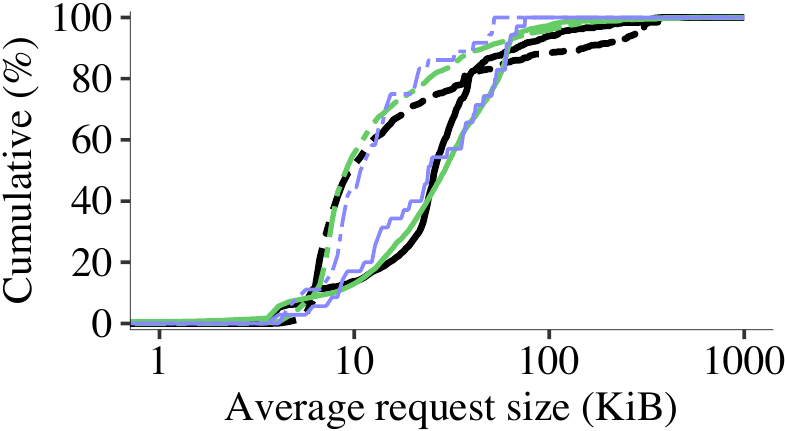} \\ 
\parbox[t]{2.2in}{\small(a) Cumulative distribution of request sizes across all
	requests} & 
\parbox[t]{2.2in}{\small(b) Cumulative distribution of average request sizes
	across all volumes} 
\vspace{6pt}\\
\includegraphics[width=2.25in]{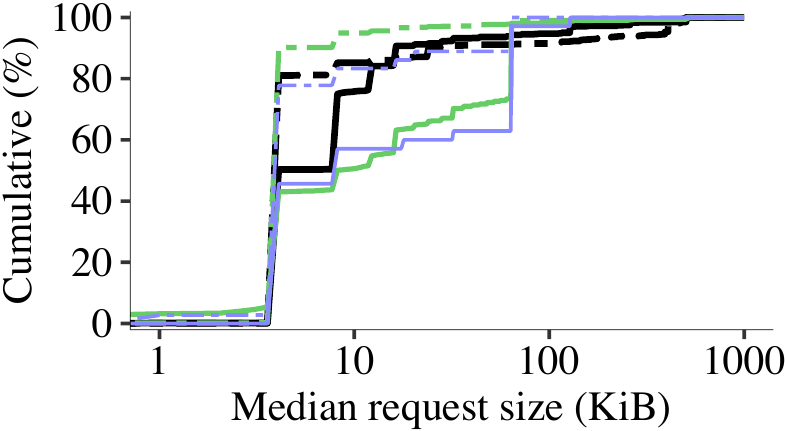} & 
\includegraphics[width=2.25in]{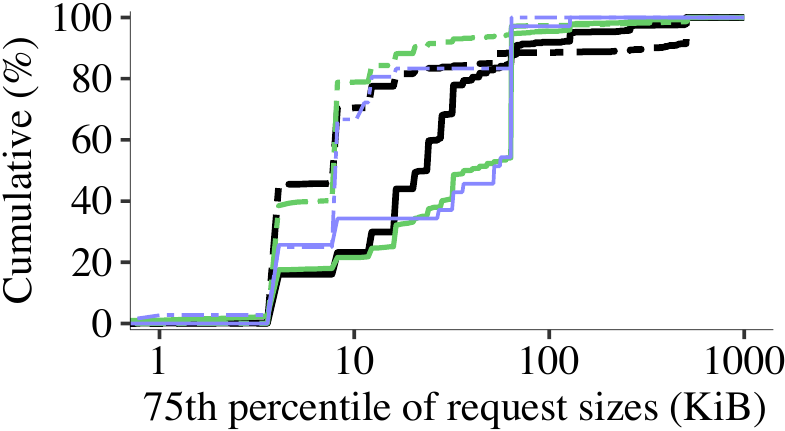} \\
\parbox[t]{2.2in}{\small(c) Cumulative distribution of median request sizes
	across all volumes} &
\parbox[t]{2.2in}{\small(d) Cumulative distribution of 75th percentiles of
        request sizes across all volumes}
\end{tabular}
\caption{Finding~A.2: Cumulative distributions of I/O request sizes.}
\label{fig:iosize}
\end{figure}

Figure~\ref{fig:iosize}(a) shows the cumulative distributions of request sizes
across all I/O requests in all three traces. We see that all traces feature
small-size I/O requests (less than 100\,KiB).  Specifically, in AliCloud, 75\%
of reads and writes are no larger than 12\,KiB and 16\,KiB, respectively,
while in TencentCloud, 75\% of reads and writes are no larger than 32\,KiB
and 12\,KiB, respectively. In MSRC, 75\% of reads and writes are no larger
than 64\,KiB and 20\,KiB, respectively. 

The dominance of small-size I/O requests also holds in individual volumes.  We
compute the average request size for each volume.  Figure~\ref{fig:iosize}(b)
shows the cumulative distributions of the average request sizes of all volumes
in all three traces. We see that 75\% of the average read and write sizes in
AliCloud are less than 37.0\,KiB and 26.8\,KiB, respectively, while 75\% of the
average read and write sizes in TencentCloud are less than 49.8\,KiB and
19.0\,KiB, respectively. For MSRC, 75\% of the average read and write sizes are
less than 48.4\,KiB and 16.4\,KiB, respectively.  Small I/Os are also commonly
found in enterprise and desktop file system workloads \cite{riska06,ahmad07}.

To ensure that our observations are not biased by outliers, we
further measure the cumulative distributions of the medians and 75th percentiles 
of request sizes across all volumes, as shown in Figures~\ref{fig:iosize}(c)
and \ref{fig:iosize}(d), respectively.  In Figure~\ref{fig:iosize}(c), we see
that the median read and write sizes in 75\% of volumes in all three traces are
no more than 64\,KiB and 4\,KiB, respectively.  Similarly, in
Figure~\ref{fig:iosize}(d), we see that the 75th percentiles of read and write
sizes in 75\% of volumes in all three traces are no more than 64\,KiB and
12\,KiB, respectively. The results indicate that small-size I/O requests
dominate even if we consider medians and 75th percentiles of (instead
of average) request sizes. 

\paragraph{Finding~A.3:} {\em A non-negligible fraction of volumes in AliCloud
are active in short time periods, but it is not the case in TencentCloud and
MSRC.}

\begin{figure}[t]
\centering
\begin{tabular}{@{\ }c@{\hspace{0.2in}}c}
\parbox[t]{2.4in}{
\includegraphics[width=2.25in]{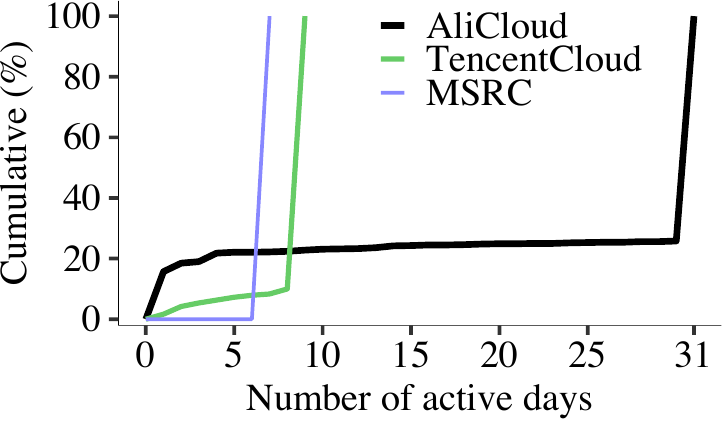} 
\caption{Finding~A.3: Cumulative distributions of numbers of active days
across all volumes.}
\label{fig:activeday}
}
&
\parbox[t]{2.4in}{
\includegraphics[width=2.25in]{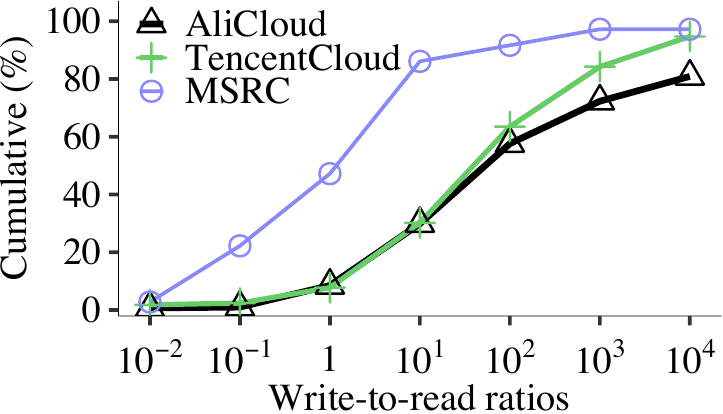} 
\caption{Finding~A.4: Cumulative distributions of write-to-read ratios across
all volumes.}
\label{fig:wr_ratio}
}
\end{tabular}
\end{figure}

We study the activeness of individual volumes. Here, we measure the
number of active days for each volume, in which a volume is said to be active
in a day if it receives at least one I/O request (i.e., up to 31, 9, and 7
active days in AliCloud, TencentCloud, and MSRC, respectively).
Figure~\ref{fig:activeday}
depicts the cumulative distributions of numbers of active days across all
volumes in all three traces. In AliCloud, 15.7\% of volumes (i.e., 157
volumes) are active for only one day. We find that 147 out of
the 157 volumes are active in only four hours, and the total WSS and I/O
traffic of the 157 volumes account for only 1.4\% and 0.07\% of all 1,000
volumes, respectively. 
One possible reason for the short active periods in such volumes
in AliCloud is the presence of short-lived tasks in cloud applications
\cite{mishra10}.  On the other hand, in TencentCloud, only 1.7\% of volumes are
active for only one day, while 90.1\% of volumes are active for all nine days
in the entire trace duration.  Also, all volumes in MSRC are active for all
seven days in the entire trace duration. 

\begin{figure}[!t] 
\centering
\begin{tabular}{@{\ }ccc}
\includegraphics[width=2in]{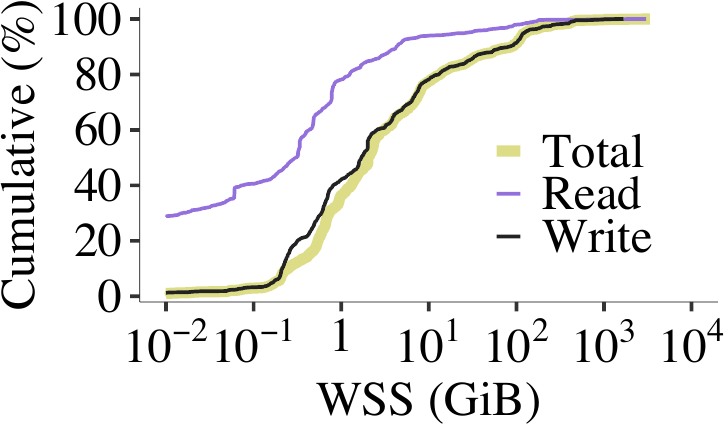} &
\includegraphics[width=2in]{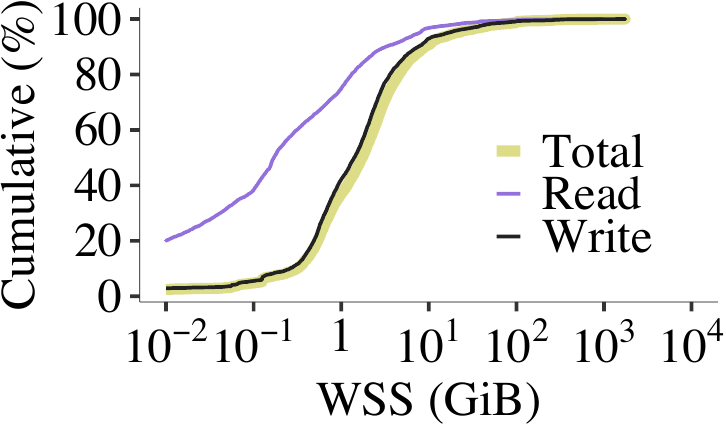} &
\includegraphics[width=2in]{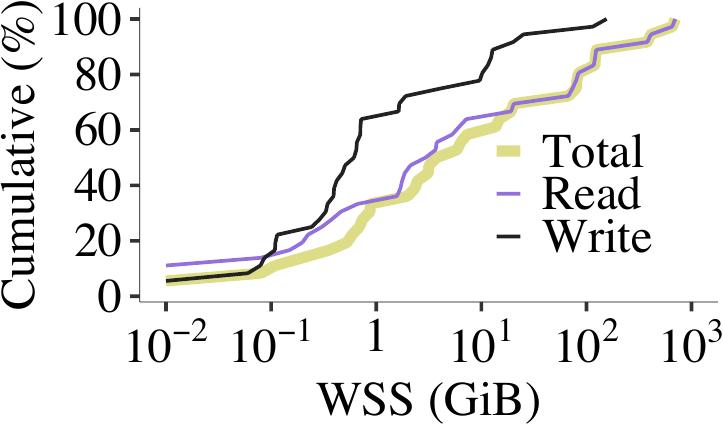} \\
{\small(a) WSSs in AliCloud} &
{\small(b) WSSs in TencentCloud} &
{\small(c) WSSs in MSRC} 
\vspace{6pt}\\
\includegraphics[width=2in]{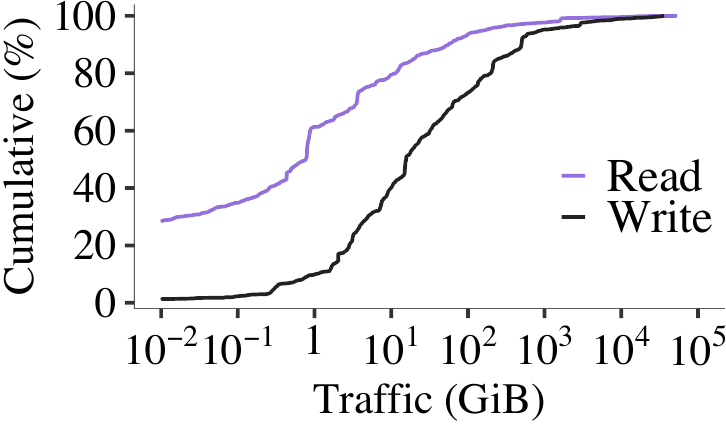} &
\includegraphics[width=2in]{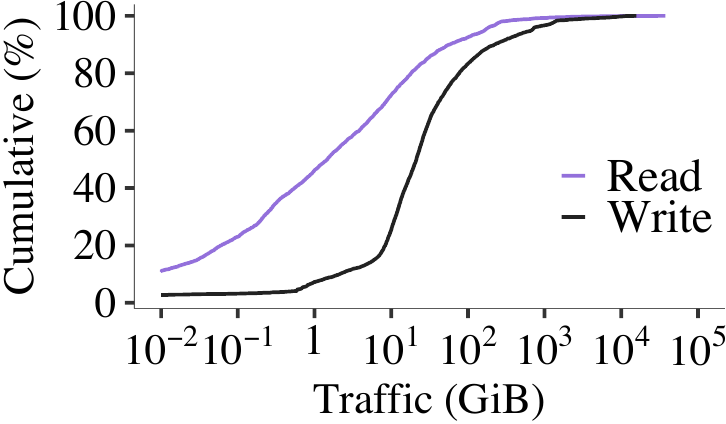} &
\includegraphics[width=2in]{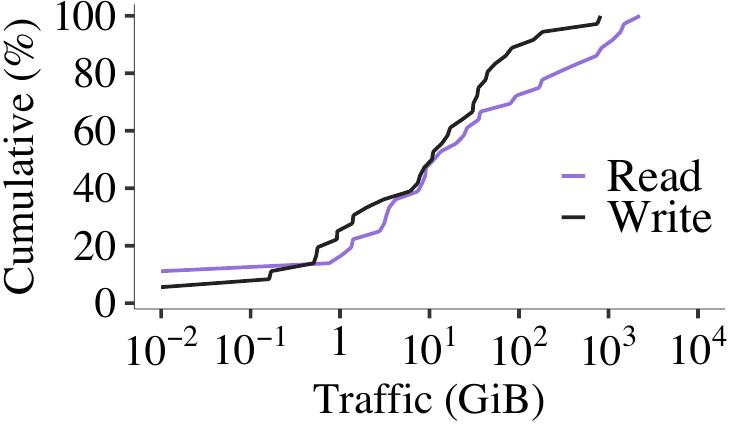} \\
{\small(d) I/O traffic sizes in AliCloud} &
{\small(e) I/O traffic sizes in TencentCloud} &
{\small(f) I/O traffic sizes in MSRC}
\end{tabular}
\vspace{-3pt}
\caption{Finding~A.4: Cumulative distributions of WSSs (figures~(a)-(c)) and
I/O traffic sizes (figures~(d)-(f)) across all volumes.}
\label{fig:wss_traffic}
\end{figure}

\paragraph{Finding~A.4:} {\em Most volumes in AliCloud and TencentCloud are
write-dominant.} 

Referring to Table~\ref{tab:basic}, the overall write-to-read ratio (i.e., the
ratio between the number of writes and the number of reads) in AliCloud is
3:1, and that in TencentCloud is 2.35:1.  However, the write-to-read ratio in
MSRC is 0.42:1 only.  We further analyze the write-to-read ratios on a
per-volume basis.  Figure~\ref{fig:wr_ratio} shows the cumulative
distributions of write-to-read ratios across all volumes in all three traces. 
In AliCloud and TencentCloud, 91.5\% (i.e., 915 out of 1,000) and 92.3\%
(i.e., 4,608 out of 4,995) of the volumes are write-dominant (i.e., the
write-to-read ratios are larger than 1).  Also, 42.4\% and 36.5\% of the
volumes in AliCloud and TencentCloud even have very high write-to-read ratios
that are larger than 100, respectively.  On the other hand, MSRC has an
opposite pattern, in which only 52.8\% (19 out of 36) of volumes are
write-dominant.  Note that prior work \cite{soundararajan10} also shows
the existence of write-dominant workloads in MSRC, especially in files such as
mail boxes, search indexes, registry files, and file system metadata files.

Figure~\ref{fig:wss_traffic} further analyzes the cumulative distributions of
WSSs and I/O traffic sizes across all volumes in all three traces.
Figures~\ref{fig:wss_traffic}(a)-\ref{fig:wss_traffic}(c) show the cumulative
distributions of the total WSSs, read WSSs, and write WSSs across all volumes. 
The write WSSs of both AliCloud and TencentCloud are significantly larger than
the read WSSs and are close to the total WSSs
(Figures~\ref{fig:wss_traffic}(a) and \ref{fig:wss_traffic}(b)).  This implies
that the total WSSs of both traces are mainly determined by writes. In
contrast, the read WSSs of MSRC are close to the total WSSs
(Figure~\ref{fig:wss_traffic}(c)).  We also make similar observations in the
I/O traffic sizes, in which AliCloud and TencentCloud are write-dominant
(Figures~\ref{fig:wss_traffic}(d)-\ref{fig:wss_traffic}(f)). 
One possible reason of the write dominance in both AliCloud and TencentCloud
is the wide use of application-level read caches in cloud storage, in which
reads are mostly absorbed in the application layer without being issued to the
storage layer \cite{wang20}.

\paragraph{Finding~A.5:} {\em In AliCloud, larger volumes tend to have 
larger percentages of the total WSS over the raw capacity.}

We examine the relationship between the total WSS and the raw capacity of a
volume.  Our analysis here focuses on the AliCloud traces, which provide the
capacity information of individual volumes.   Specifically, we divide the
1,000 volumes in AliCloud into four groups by volume capacities, including
40-49\,GiB, 50-99\,GiB, 100-199\,GiB, and 200-5,000\,GiB (note that each
volume capacity is represented as an integer GiB). They account for 444, 179,
170, and 207 volumes, respectively.  For each volume, we calculate the
WSS-to-capacity percentage (i.e., the percentage of the total WSS over the raw
capacity of the volume).  We then plot the cumulative distributions of the
WSS-to-capacity percentages for the four groups. 

\begin{figure}[t]
\centering
\begin{tabular}{@{\ }c@{\hspace{0.2in}}c}
\parbox[t]{2.4in}{
\includegraphics[width=2.25in]{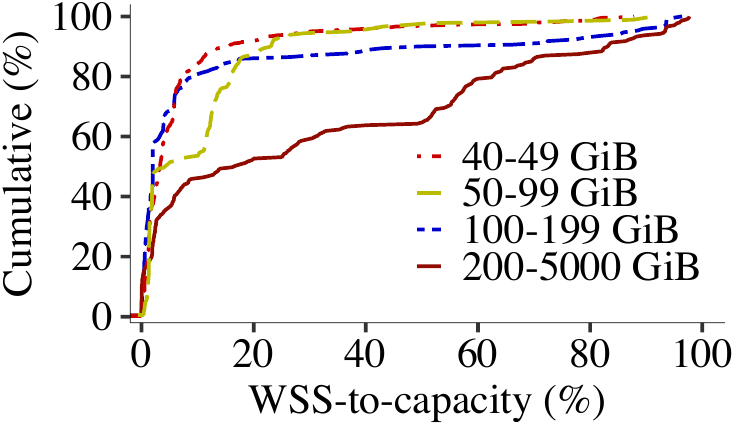} 
\caption{Finding~A.5: Cumulative distributions of WSS-to-capacity
percentages across all volumes of different groups of raw capacities in
AliCloud.}
\label{fig:cap_wss}
}
&
\parbox[t]{2.4in}{
\includegraphics[width=2.25in]{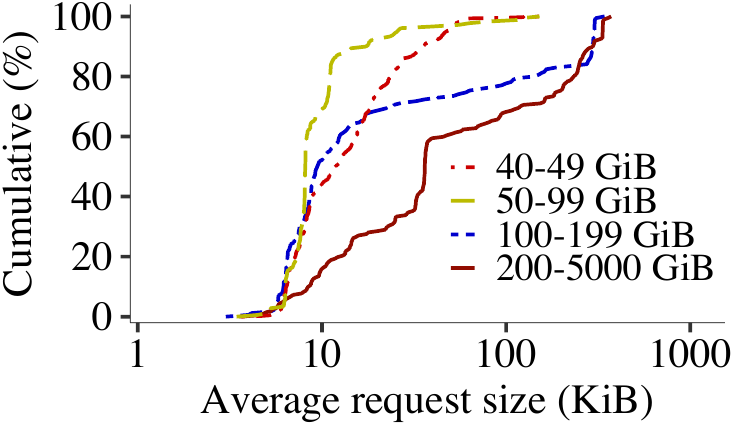} 
\caption{Finding~A.6: Cumulative distributions of average request
sizes across all volumes of different groups of raw capacities in AliCloud.}
\label{fig:cap_reqsize}
}
\end{tabular}
\end{figure}

Figure~\ref{fig:cap_wss} shows the results.  For small volumes, the
WSS-to-capacity percentages tend to be low (i.e., the usage of disk space is
low).  Specifically, 80\% of the volumes in the 40-49\,GiB, 50-99\,GiB, and 
100-199\,GiB groups have WSS-to-capacity percentages of less than 7.3\%,
16.1\%, and 9.7\%, respectively.  On the other hand, in the 200-5,000\,GiB
group, half of the volumes have a WSS-to-capacity percentage of more than
16.7\%, and 35.7\% of the volumes have WSS-to-capacity percentages of more than
50\%.  In particular, the largest volume (i.e., with a raw capacity of
5,000\,GiB) has a WSS-to-capacity percentage of 66.9\%. 

\paragraph{Finding~A.6:} {\em In AliCloud, larger volumes tend to
have larger write request sizes, but have similar read request sizes compared
to smaller volumes. } 

We examine the average request sizes of individual volumes with respect to
their raw capacities.  We again divide the volumes by raw capacities into four
groups (i.e., 40-49\,GiB, 50-99\,GiB, 100-199\,GiB, and 200-5,000\,GiB) as
above. 

Figure~\ref{fig:cap_reqsize} shows the results. Overall, the average request
sizes in the 200-5,000\,GiB group are larger than in the other three groups
with smaller raw capacities.  Specifically, the median of average request
sizes in the 200-5,000\,GiB group is 36.2\,KiB, while the medians of average
request sizes in the 40-49\,GiB, 50-99\,GiB, and 100-199\,GiB groups are
11.9\,KiB, 8.1\,KiB, and 9.5\,KiB, respectively.  A possible reason is that
larger volumes are often related to the workloads with the larger-size
sequential data accesses. 

\begin{figure}[!t] 
\centering
\begin{tabular}{@{\ }cc}
\includegraphics[width=2.25in]{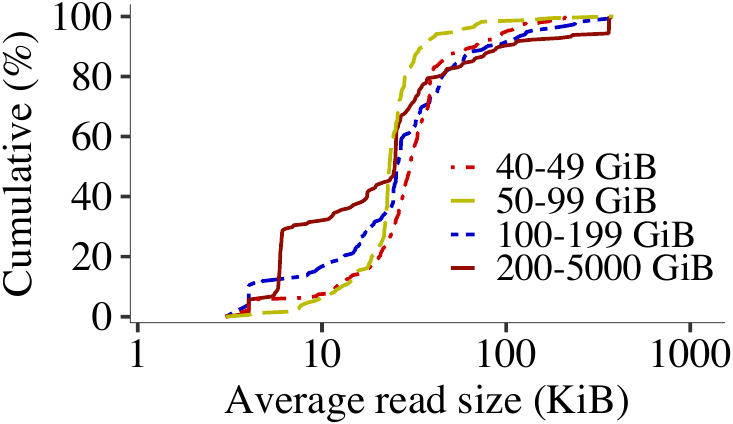} &
\includegraphics[width=2.25in]{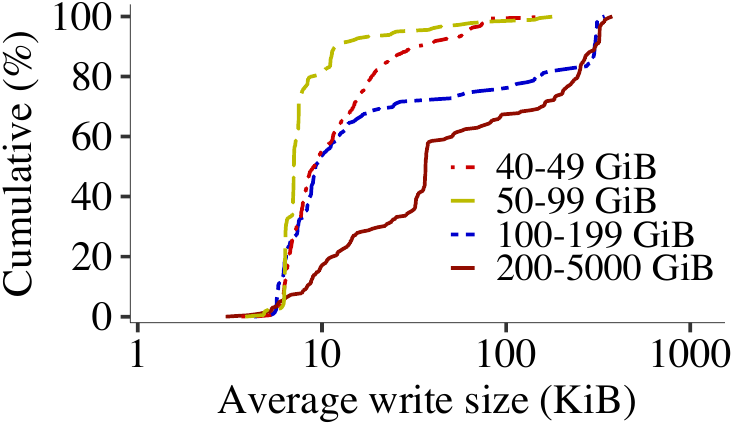} \\ 
{\small (a) Average read request sizes} & 
{\small (b) Average write request sizes}  
\end{tabular}
\vspace{-3pt}
\caption{Finding~A.6: Cumulative distributions of average read and write
request sizes across all volumes of different groups of raw capacities in
AliCloud.}
\label{fig:cap_reqsize_rw}
\end{figure}

We further examine the average read and write request sizes of
individual volumes to understand where the larger requests in larger volumes
come from. Figure~\ref{fig:cap_reqsize_rw} shows the results. We find that the
average read sizes are close in four groups; the medians of average read
request sizes in the 40-49\,GiB, 50-99\,GiB, 100-199\,GiB, and 200-5,000\,GiB
groups are 29.3\,KiB, 23.0\,KiB, 24.9\,KiB, and 24.4\,KiB, respectively
(Figure~\ref{fig:cap_reqsize_rw}(a)).  In contrast, the median of average
write request sizes in the 200-5000\,GiB group is 36.5\,KiB, while the medians
of average write request sizes in the 40-49\,GiB, 50-99\,GiB, and 100-199\,GiB
groups are 9.1\,KiB, 7.0\,KiB, and 9.2\,KiB, respectively
(Figure~\ref{fig:cap_reqsize_rw}(b)).  

\paragraph{Summary.} AliCloud, TencentCloud, and MSRC have some common
aspects, such as the dominance of small-size I/O requests, yet they also have
many differences. In particular, both AliCloud and TencentCloud are
write-dominant, while MSRC is read-dominant.  A unique aspect for AliCloud, as
opposed to TencentCloud and MSRC, is that it has a non-negligible fraction of
volumes with short active periods.  Also, large volumes in AliCloud tend to
have larger WSS-to-capacity percentages and larger request sizes. 

\section{Detailed Analysis}
\label{sec:finding}

In this section, we conduct an in-depth comparative analysis on AliCloud,
TencentCloud, and MSRC in three aspects: load intensities, spatial patterns,
and temporal patterns. We report 16 findings from our analysis.
Table~\ref{tab:obsv_summary} summarizes and compares the key properties of all
three traces observed in our detailed analysis. 

\begin{table}[!t]
\small
\centering
\renewcommand{\arraystretch}{1.15}
\setlength{\tabcolsep}{4pt}
\begin{tabular}{c|c|c|c|c}
\hline
 & {\bf Property} & {\bf AliCloud} & {\bf TencentCloud} & {\bf MSRC} \\ 
\hline
\hline
\multirow{9}{*}{\makecell{{\bf Load} \\ {\bf intensity}}} &
{\bf Average intensities (B.1) } & 
\multicolumn{3}{c}{Similar} \\
\cline{2-5} & 
  {\bf Peak intensities (B.1) } & Low & Low & High \\ 
\cline{2-5} & 
{\bf Burstiness in volumes (B.2) } 
    & Yes & Yes & Yes \\
\cline{2-5} & 
{\bf Diversity of burstiness (B.3) } 
    & High & High & Low         \\ 
\cline{2-5} & 
{\bf Inter-arrival times of requests (B.4) } 
    & Short  & - & Short         \\ 
\cline{2-5} & 
{\bf Activeness (B.5) } 
    & High & Highest & Low         \\ 
\cline{2-5} & 
{\bf Activeness dominated by writes (B.6)} 
    & Yes & Yes & Yes \\
\cline{2-5} & 
{\bf Activeness of reads (B.7)} 
    & Low & Highest & High   \\ 
\cline{2-5} & 
{\bf I/O traffic in daytime (B.8)} 
    & 52.0\% & 49.6\% & 23.3\%   \\ 
\hline
\multirow{4}{*}{\makecell{{\bf Spatial} \\ {\bf patterns}}} &
{\bf Fractions of random I/Os (B.9)} 
    & High & Highest & Low   \\ 
\cline{2-5} & 
{\bf Spatial aggregations of reads (B.10 and B.11)} 
    & High & Highest & High   \\ 
\cline{2-5} & 
{\bf Spatial aggregations of writes (B.10 and B.11)} 
    & High & High & Low   \\ 
\cline{2-5} & 
{\bf Update coverages (B.12)} & High & High & Low   \\ 
\hline
\multirow{8}{*}{\makecell{{\bf Temporal} \\ {\bf patterns}}} &
{\bf Read-after-write (RAW) times (B.13)} & Large & Small & Large   \\ 
\cline{2-5} & 
{\bf Write-after-write (WAW) times (B.13)} & Large & Small & Small   \\ 
\cline{2-5} & 
{\bf More WAW requests than RAW requests (B.13)} & Yes & Yes & No \\ 
\cline{2-5} & 
{\bf Read-after-read (RAR) times (B.14)} & Large & Small & Large   \\ 
\cline{2-5} & 
{\bf Write-after-read (WAR) times (B.14)} & Large & Small & Large  \\ 
\cline{2-5} & 
{\bf More RAR requests than WAR requests (B.14)} & Yes & Yes & Yes \\ 
\cline{2-5} & 
{\bf Varying update intervals (B.15)} & Yes & Yes & Yes \\ 
\cline{2-5} & 
{\bf Miss ratios (B.16)} & High & Low & High \\ 
\hline
\end{tabular}
\vspace{3pt}
\caption{Summary of the key properties of AliCloud, TencentCloud, and MSRC in
Findings~B.1-B.16.}
\label{tab:obsv_summary}
\end{table}

\subsection{Load Intensities}
\label{subsec:intensity}

We study the characteristics of load intensities in the volumes of
AliCloud, TencentCloud, and MSRC through a number of metrics.
Specifically, we examine the average and peak load intensities
\cite{narayanan08} and the distribution of inter-arrival times of requests
\cite{wajahat19}. We also examine the activeness of volumes through the number
of active volumes \cite{narayanan08} and the active period of each volume.

\paragraph{Finding~B.1:} {\em AliCloud, TencentCloud, and MSRC have similar
average load intensities of volumes, while the peak load intensities of
AliCloud and TencentCloud are generally lower than that of MSRC.}

We measure the load intensities of individual volumes, in terms of the number
of requests per second (req/s), in two aspects. We first measure the 
{\em average intensity} of a volume, defined as the total number of requests
divided by the time elapsed between the first and last requests of the volume.
Note that for TencentCloud, if the missing hour of requests lies between the
first and last requests (\S\ref{subsec:overview}), we subtract the elapsed
time by one hour.  We also measure the {\em peak intensity} of a
volume, in which we divide the whole duration of requests of the volume into
one-minute intervals and find the peak intensity as the maximum number of
requests (per second) across all intervals; we use one-minute
intervals instead of one-second intervals since one-minute intervals are long
enough to accumulate sufficient bursty requests.

\begin{figure}[!t] 
\centering
\begin{tabular}{@{\ }ccc}
\includegraphics[width=2in]{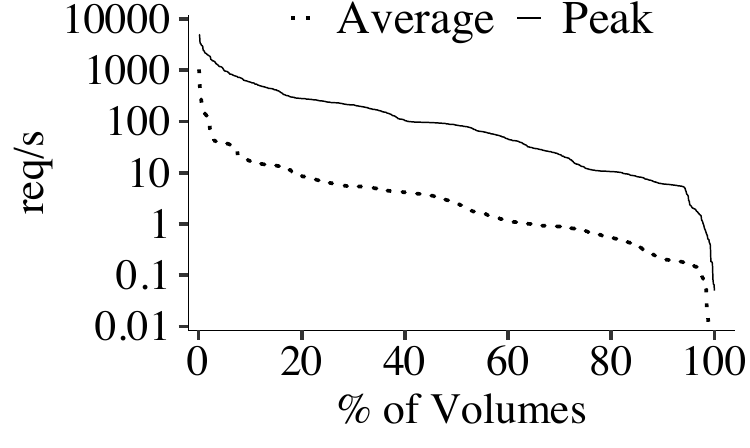} &
\includegraphics[width=2in]{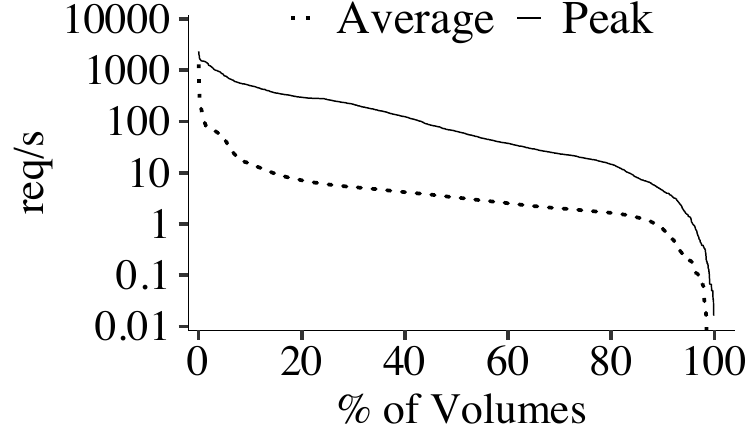} &
\includegraphics[width=2in]{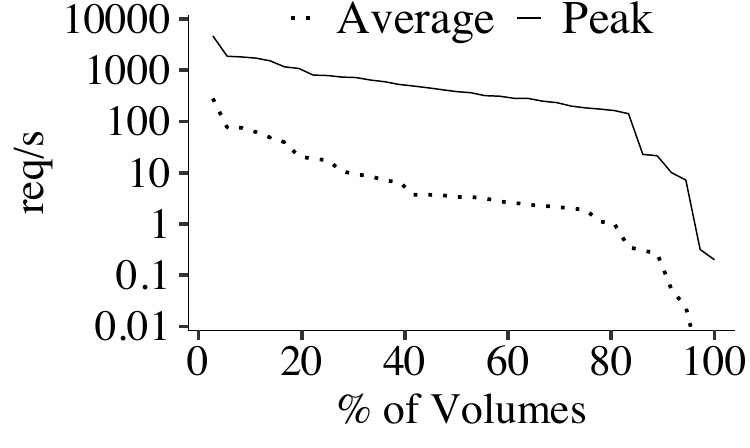} \\
{\small (a) AliCloud} & 
{\small (b) TencentCloud} & 
{\small (c) MSRC} 
\end{tabular}
\vspace{-3pt}
\caption{Finding~B.1: Average and peak intensities of volumes.  Note that we
sort the average and peak intensities of volumes in descending order in their
respective curves to make the distribution of each type of intensities clearly
shown, so the curves are different from the ones in our conference version
\cite{li20}.}
\label{fig:peak}
\end{figure}

Figure~\ref{fig:peak} shows the average and peak intensities of volumes in
AliCloud, TencentCloud, and MSRC, sorted by the average intensities of volumes
in descending order. We observe similar trends of average intensities in all
three traces, but different patterns of peak intensities in TencentCloud. 
In AliCloud, TencentCloud, and MSRC,
only 1.90\%, 1.16\%, and 2.78\% of volumes have average intensities above
100\,req/s, and the percentages of volumes with average intensities lower than
10\,req/s are 81.6\%, 85.7\%, and 72.2\%, respectively.  Furthermore, their
medians of average intensities are 2.55\,req/s, 3.27\,req/s, and 3.36\,req/s,
respectively. A possible reason of having similar average intensities in all
three traces is that more applications are moving to the cloud \cite{li19}, so
the average intensities are similar in both cloud and traditional data center
environments. However, as for the peak intensities, their 90th percentiles of
peak intensities are 578.7\,req/s, 498.9\,req/s, and 1,612.1\,req/s,
respectively, in which the peak intensity of MSRC is much higher than those of
other two traces. 

\begin{figure}[!t]
  \begin{minipage}[b]{2.6in}
    \centering
    \includegraphics[width=2.2in]{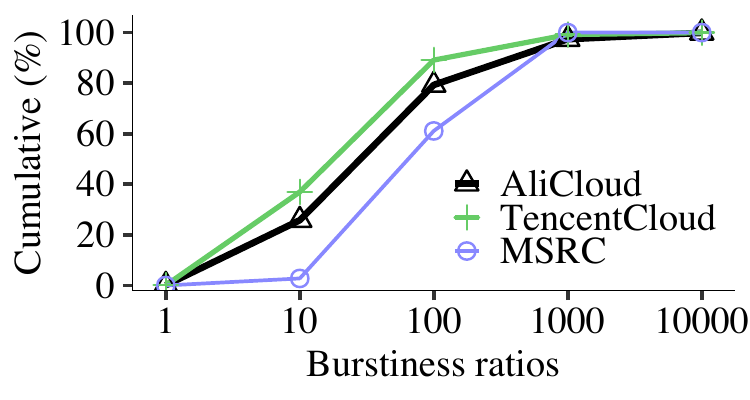} 
    \vspace{-6pt}
    \captionof{figure}{Findings~B.2-B.3: Cumulative distribution of burstiness ratios of
    volumes.}
    \label{fig:burst}
  \end{minipage}
  \hspace{0.1in}
  \begin{minipage}[b]{3.3in} \small 
  \centering
  \captionsetup{type=table}
  \renewcommand{\arraystretch}{1.2}
  \vspace{10pt}
  \begin{tabular}{c|c|c|c}
  \hline
  Traces          &  AliCloud        & TencentCloud & MSRC \\ 
  \hline
  \hline
  Peak intensity (req/s)    &  15,965.8     & 70,910.1 & 5,296.8         \\ 
  \hline
  Average intensity (req/s) &  7,554.1     & 43,036.0  & 717.2          \\ 
  \hline
  Burstiness ratio          &  2.11     & 1.65    & 7.39           \\ 
  \hline
  \end{tabular}
  \vspace{10pt}
  \captionof{table}{Finding~B.2: Overall peak and average intensities as well as burstiness
  ratios.}
  \label{tab:burst}
  \vspace{-2pt}
  \end{minipage}
\end{figure}

\paragraph{Finding~B.2:} {\em AliCloud, TencentCloud, and MSRC have high
burstiness in a non-negligible fraction of volumes, but their overall
burstiness is mild.}
	
We examine the burstiness of all three traces by measuring the {\em burstiness
ratio} of a volume, defined as the ratio between the peak intensity and the
average intensity of the volume.  Figure~\ref{fig:burst} shows the cumulative
distributions of burstiness ratios across all volumes in AliCloud,
TencentCloud, and MSRC.  We see that a non-negligible fraction of volumes
(20.7\%, 10.9\%, and 38.9\% in AliCloud, TencentCloud, and MSRC, respectively)
have burstiness ratios higher than 100.  Also, 74.2\%, 63.0\%, and 97.2\% of
the volumes in AliCloud, TencentCloud, and MSRC have burstiness ratios higher
than 10, respectively.  This implies that burstiness is common, and such
bursty volumes can observe load imbalance at some time.  Note that
prior work \cite{riska06} also shows that burstiness exists across various
types of applications, such as enterprise systems, desktops, consumer
electronics, web servers, and file systems.  On the other hand, if we examine
overall burstiness level by aggregating all volumes of the whole traces, the
burstiness ratios are mild, with 2.11 in AliCloud, 1.65 in TencentCloud, and
7.39 in MSRC (see Table~\ref{tab:burst}).  Compared with both AliCloud and
MSRC, TencentCloud has a lower percentage of volumes with high burstiness
ratios, as well as a lower overall burstiness.  This shows that the burstiness
level is mild from the whole-system's perspective, but is significant for some
of the volumes.

\paragraph{Finding~B.3:} {\em AliCloud and TencentCloud have more
diverse burstiness across volumes than MSRC.}

The volumes in both AliCloud and TencentCloud span a wider range of
burstiness than those in MSRC.  Referring to Figure~\ref{fig:burst}, for the
volumes with low burstiness, 25.8\% and 37.0\% of volumes in
AliCloud and TencentCloud have burstiness ratios less than 10,
while the corresponding percentage is only 2.78\% in MSRC. On the
other hand, for the volumes with high burstiness, 2.60\% and 0.80\%
of volumes in AliCloud and TencentCloud have burstiness ratios larger than
1,000, respectively, while there are no such volumes in MSRC.  The higher
diversities of burstiness in AliCloud and TencentCloud suggest larger
variations in workload characteristics among different volumes in cloud block
storage.

\paragraph{Finding~B.4:} {\em Both AliCloud and MSRC have high short-term
burstiness from the perspective of inter-arrival times of requests.}

\begin{figure}[!t] 
\centering
\begin{tabular}{@{\ }ccc}
\includegraphics[width=2in]{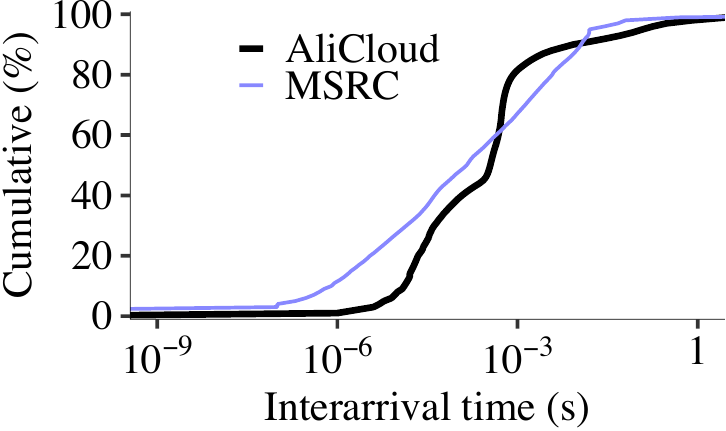} &
\includegraphics[width=2in]{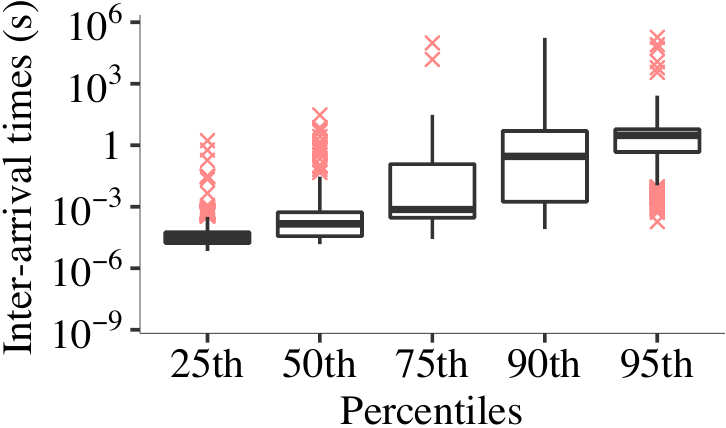} &
\includegraphics[width=2in]{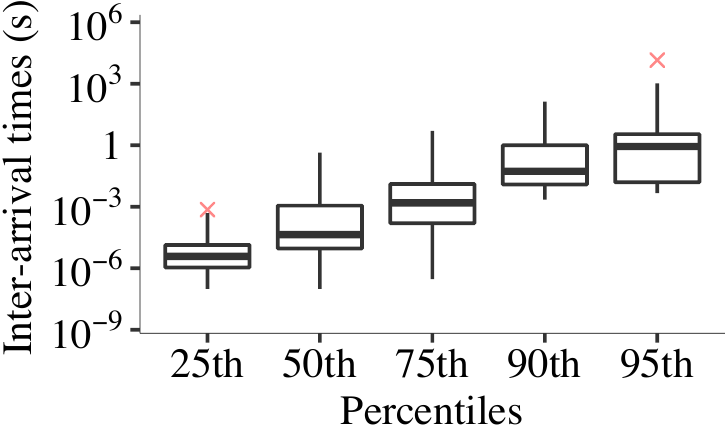} \\
{\small (a) Cumulative distributions} & 
{\small (b) AliCloud}  & 
{\small (c) MSRC} 
\end{tabular}
\vspace{-3pt}
\caption{Finding~B.4: Inter-arrival times of requests. Figure~(a)
shows the cumulative distributions of inter-arrival times across all volumes.
In figures~(b) and (c), each boxplot represents the distribution of all the
values collected in each volume according to the corresponding percentile.  } 
\label{fig:inter}
\end{figure}

We measure the inter-arrival times of I/O requests (i.e., the elapsed time
between two adjacent requests) for each volume.  We first examine
the cumulative distributions of inter-arrival times across all volumes.
Figure~\ref{fig:inter}(a) shows the results.  Note that we do not consider
TencentCloud, as its timestamps are in units of seconds and we cannot
accurately measure the inter-arrival times at finer-grained granularities
(e.g., at the microsecond level). We find that most of the inter-arrival times
are smaller than one second. For example, in AliCloud and MSRC, 50\% of the
inter-arrival times are smaller than 351\,$\mu$s and 142\,$\mu$s,
respectively, and 99\% of the inter-arrival times are smaller than 3,140\,ms
and 484\,ms, respectively.  The results indicate that short inter-arrival
times are common in both traces.

We also consider five groups of percentiles of inter-arrival times for each
volume, including the 25th, 50th, 75th, 90th, and 95th percentiles.  We
represent each group of percentile values of all volumes by boxplots.
Figures~\ref{fig:inter}(b) and \ref{fig:inter}(c) show the results of both
AliCloud and MSRC, respectively.
Both AliCloud and MSRC traces have a high number of bursty requests,
as indicated by large fractions of short inter-arrival times in the volumes.
In particular, the medians of the groups of 25th, 50th, and 75th percentiles
are lower than 1.3\,ms, or equivalently over 700\,req/s (i.e., 31\,$\mu$s,
145\,$\mu$s, and 735\,$\mu$s in AliCloud, and 3.5\,$\mu$s, 30.5\,$\mu$s, and
1.3\,ms in MSRC, respectively).  Also, the volumes in AliCloud have much higher
inter-arrival times of requests than those in MSRC.  For example, half of the
volumes in AliCloud have 25th percentiles higher than 31\,$\mu$s
(Figure~\ref{fig:inter}(a)), while half of the volumes in MSRC have 25th
percentiles higher than 3.5\,$\mu$s (Figure~\ref{fig:inter}(b)). 
Note that prior work \cite{kavalanekar08} also identifies the existence of
short inter-arrival times (e.g., a few milliseconds) in server workloads.

\paragraph{Finding~B.5:} {\em Most of the volumes in AliCloud, TencentCloud,
and MSRC are active throughout the trace periods, while AliCloud and
TencentCloud are more active than MSRC. TencentCloud has the highest
activeness among all three traces, from both the perspectives of active
volumes and active time periods.}
	
Recall from Section~\ref{subsec:highlevel} that we examine the activeness of
volumes of all three traces on a per-day basis.  We now revisit the activeness
of volumes of all three traces in a more fine-grained
manner.  Specifically, we divide the traces into 10-minute intervals. We say
that a volume is {\em active} in an interval if it has at least one request in
the interval.  We also say that a volume is {\em read-active} and {\em
write-active} in an interval if it has at least one read request and one write
request in the interval, respectively. 

\begin{figure}[!t] 
\centering
\begin{tabular}{@{\ }ccc}
\multicolumn{3}{c}{\includegraphics[width=2.4in]{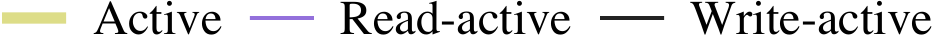}} \\
\includegraphics[width=2in]{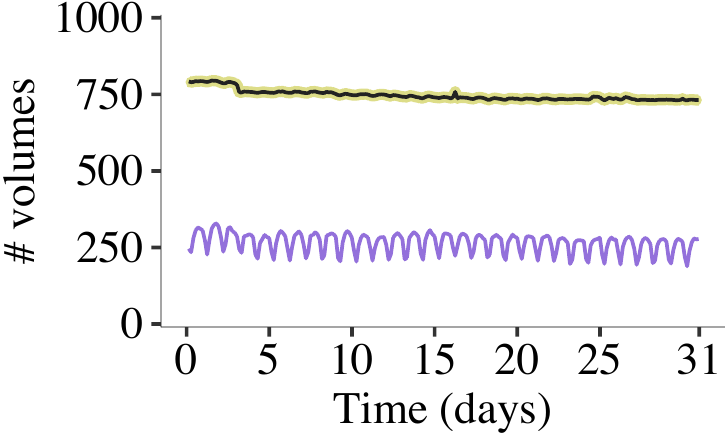} &
\includegraphics[width=2in]{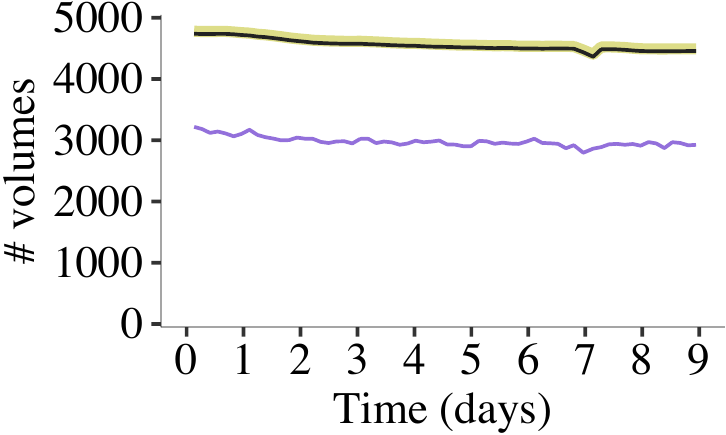} &
\includegraphics[width=2in]{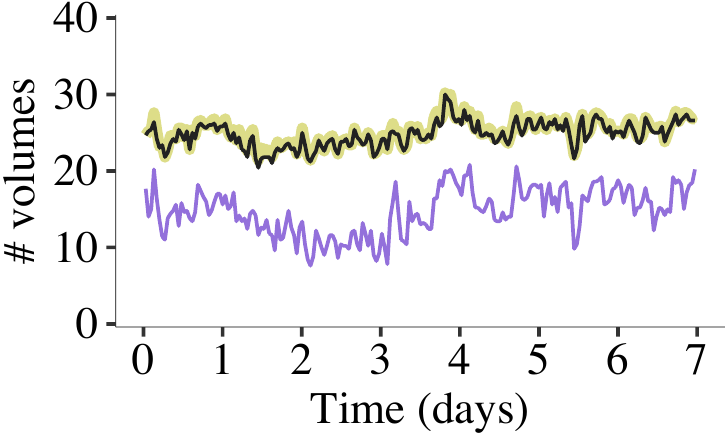} \\
{\small (a) AliCloud} & 
{\small (b) TencentCloud} & 
{\small (c) MSRC}
\end{tabular}
\vspace{-3pt}
\caption{Findings~B.5-B.7: Numbers of active, read-active, and write-active
volumes. Note that the ``Active'' and ``Write-active'' curves almost overlap
with each other.}
\label{fig:active}
\end{figure}

Figure~\ref{fig:active} depicts the numbers of active, read-active, and
write-active volumes throughout the trace periods in AliCloud, TencentCloud,
and MSRC (recall that they have 1,000, 4,995, and 36 volumes,
respectively).  We find that the percentages of active volumes throughout the
trace duration are always larger than 73.1\%, 88.2\%, and 59.4\% in AliCloud,
TencentCloud, and MSRC, respectively. In particular, TencentCloud has the
highest fraction of active volumes throughout the trace periods. Also, the
numbers of active volumes in AliCloud and TencentCloud have more stable trends
compared with that
in MSRC. Furthermore, the number of read-active volumes in AliCloud shows
diurnal patterns by often having less than 200 read-active volumes at night
and more than 300 read-active volumes at daytime, which is similar to the
patterns found in object storage systems \cite{beaver10}, enterprise virtual
desktops \cite{lee17}, and key-value stores \cite{atikoglu12, yang20, cao20}.

\begin{figure}[!t] 
\centering
\begin{tabular}{@{\ }ccc}
\multicolumn{3}{c}{\includegraphics[width=2.4in]{figs/b5_b6_b7/b5_b6_b7_time_legend.pdf}} \\
\includegraphics[width=2in]{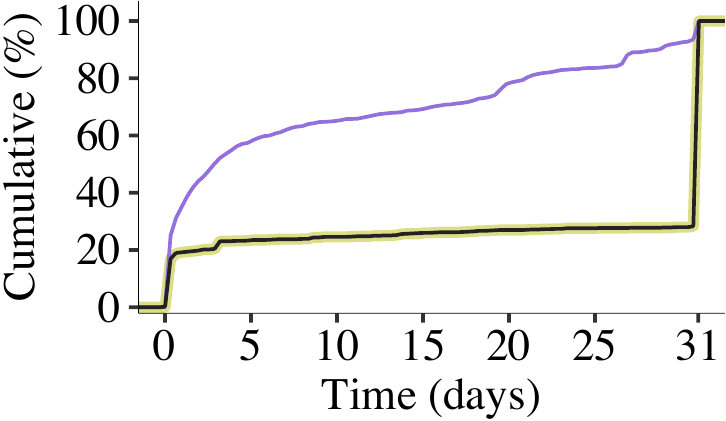} &
\includegraphics[width=2in]{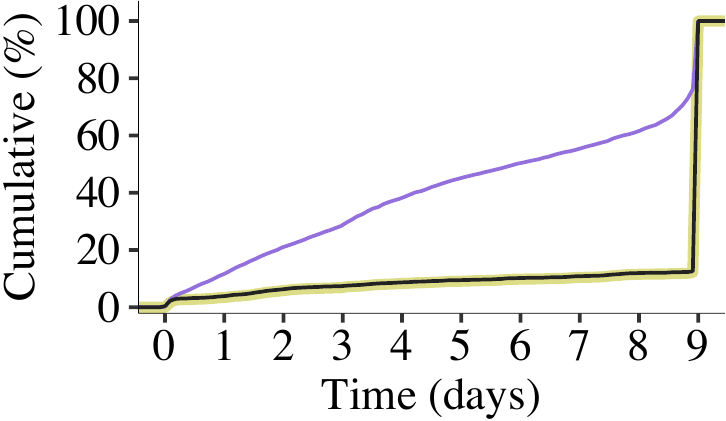} &
\includegraphics[width=2in]{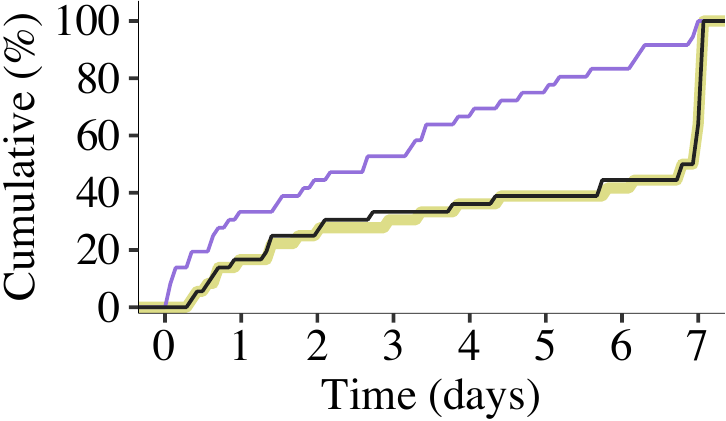} \\
{\small (a) AliCloud}  & 
{\small (b) TencentCloud}  & 
{\small (c) MSRC} 
\end{tabular}
\vspace{-3pt}
\caption{Findings~B.5-B.7: Cumulative distributions of active time periods
measured in 10-minute intervals across all volumes.  Note that the ``Active''
and ``Write-active'' curves almost overlap with each other.}
\label{fig:active_time}
\end{figure}

We also measure the active time period of each volume, based on the number of
10-minute intervals in which the volume is active.
Figure~\ref{fig:active_time} depicts the cumulative percentages of active time
periods across all volumes in all three traces. More than 72.2\%, 88.2\%, and
55.6\% of the volumes are active during 95\% of the whole trace periods in
AliCloud, TencentCloud, and MSRC, respectively. This indicates that most of
the volumes in all three traces have high activeness throughout the whole
trace periods, and AliCloud and TencentCloud have higher activeness in general
than MSRC.  Also, in terms of the active time in each volume, the activeness
is the highest in the TencentCloud volumes.

\paragraph{Finding~B.6:} {\em Writes are the dominant factor in determining
the activeness in AliCloud, TencentCloud, and MSRC.} 

Referring to both Figures~\ref{fig:active} and \ref{fig:active_time}, the
curves of ``Active'' and ``Write-active'' nearly overlap with each other in
all three traces. It suggests that the activeness of all three
traces (in terms of the number of active volumes and the active time period of
a volume) is mainly determined by the presence of writes. Thus, load balancing
on writes is important for the volumes in cloud block storage. 

\paragraph{Finding~B.7:} {\em Removing write requests shows drastic decreases
in activeness in AliCloud, TencentCloud, and MSRC. AliCloud is less
read-active than MSRC, and TencentCloud is the most read-active among all
three traces.}

If we remove write requests and consider only the read-active volumes,
Figure~\ref{fig:active} shows that the number of active volumes decreases
drastically.  In particular, the number of active volumes reduces by
58.6-74.2\% in AliCloud (Figure~\ref{fig:active}(a)), 32.7-37.5\% in
TencentCloud (Figure~\ref{fig:active}(b)), and 24.6-65.8\% in MSRC
(Figure~\ref{fig:active}(c)).

Figure~\ref{fig:active_time} further shows that the volumes in all three traces
have low read-active time, which is consistent with results of
prior work \cite{narayanan08}. In AliCloud, TencentCloud, and
MSRC, half of the volumes have less than 2.83~days, 5.92~days, and 2.61~days of
read-active time after removing writes, corresponding to 9.1\%, 65.4\%, and
37.3\% of their whole trace durations, respectively 
(Figures~\ref{fig:active_time}(a)-\ref{fig:active_time}(c)).
In terms of the percentage of volumes that have long read-active time,
we find that 7.6\%, 38.4\%, 16.7\% of the volumes 
can reach more than 30~days, 8~days, and 6~days of read-active time in
AliCloud, TencentCloud, and MSRC, respectively (recall that their trace
durations are 31, 9.04, and 7~days, respectively). This suggests that AliCloud
is less read-active than MSRC, while TencentCloud is the most read-active among
all three traces.

\begin{figure}[!t] 
\centering
\begin{tabular}{@{\ }ccc}
\multicolumn{3}{c}{\includegraphics[width=1.6in]{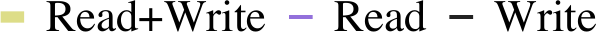}} \\
\includegraphics[width=2in]{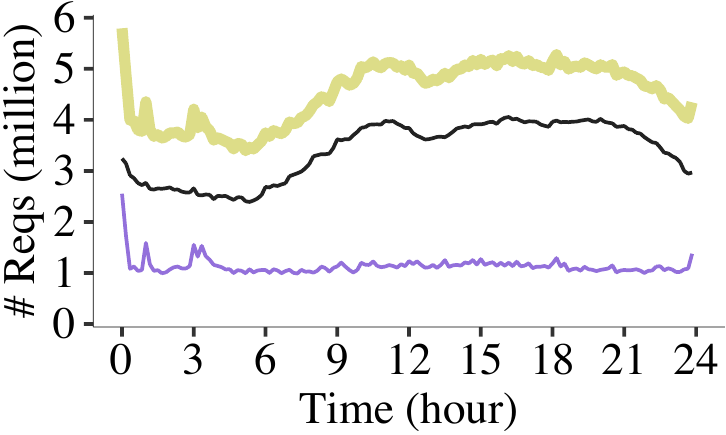} &
\includegraphics[width=2in]{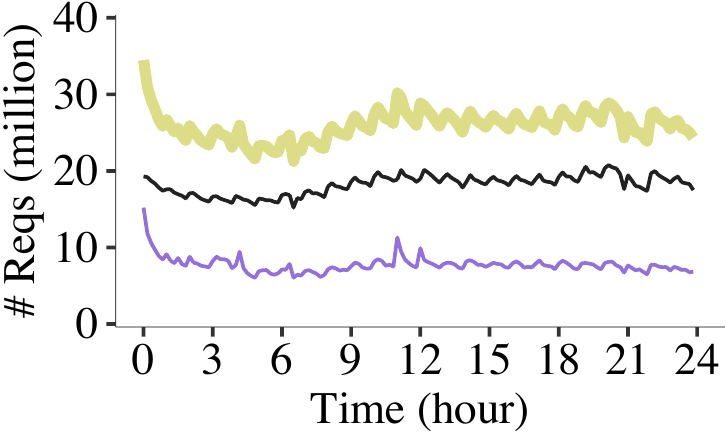} &
\includegraphics[width=2in]{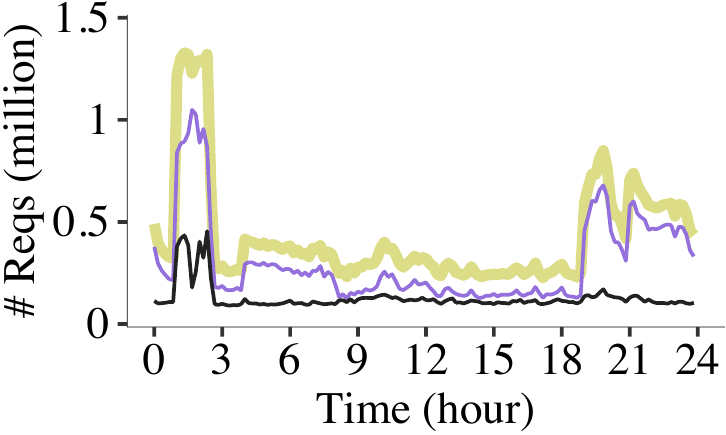} \\
{\small (a) AliCloud}  & 
{\small (b) TencentCloud}  & 
{\small (c) MSRC} 
\end{tabular}
\vspace{-3pt}
\caption{Finding~B.8: Average number of requests in 10-minute
intervals each day from 12:00~AM to 11:59~PM.}
\label{fig:per_hour_request}
\vspace{-3pt}
\end{figure}

\begin{figure}[!t] 
\centering
\begin{tabular}{@{\ }ccc}
\multicolumn{3}{c}{\includegraphics[width=1.6in]{figs/b8/b8_legend.pdf}} \\
\includegraphics[width=2in]{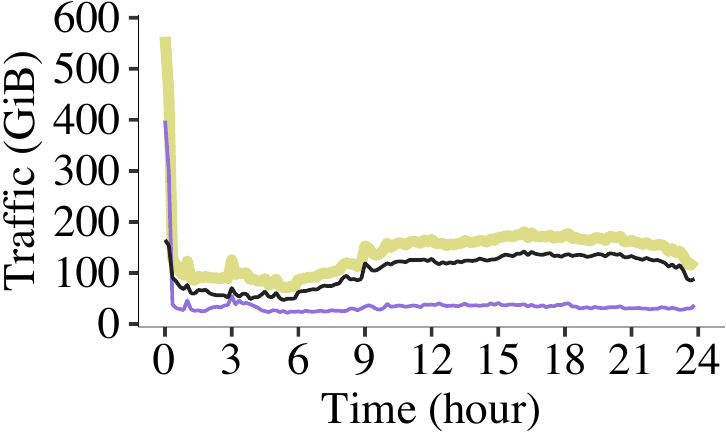} &
\includegraphics[width=2in]{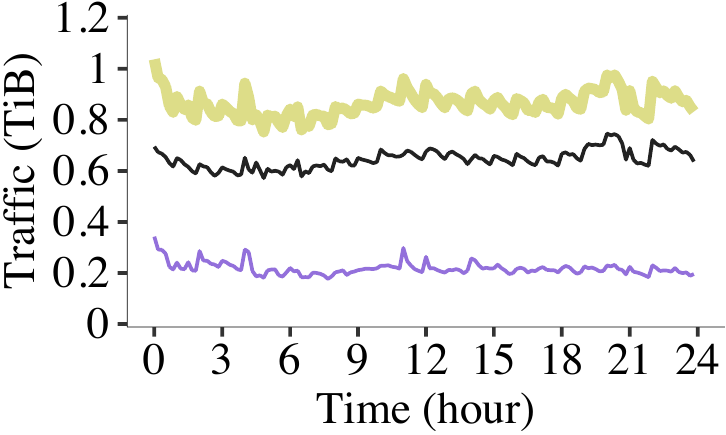} &
\includegraphics[width=2in]{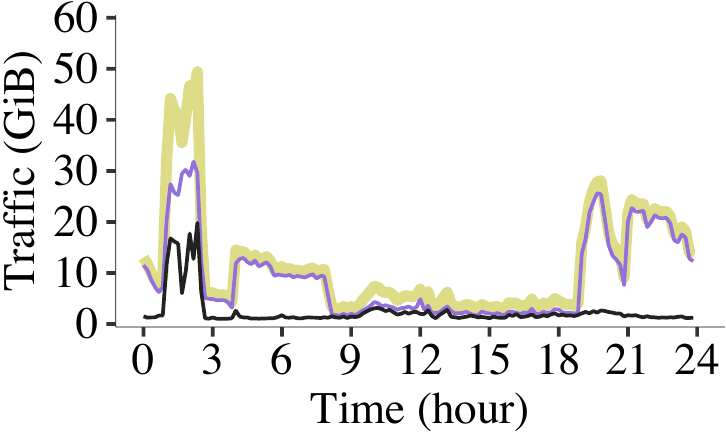} \\
{\small (a) AliCloud}  & 
{\small (b) TencentCloud}  & 
{\small (c) MSRC} 
\end{tabular}
\vspace{-3pt}
\caption{Finding~B.8: Average amount of traffic in 10-minute
intervals each day from 12:00~AM to 11:59~PM. }
\label{fig:per_hour_traffic}
\vspace{-3pt}
\end{figure}

\paragraph{Finding~B.8: } {\em I/O traffic in both AliCloud and
TencentCloud is almost evenly spread across daytime and nighttime.  AliCloud
and MSRC have large amounts of read traffic near midnight.}

We calculate the average number of requests and the average amount of traffic
in 10-minute intervals each day from 12:00~AM to 11:59~PM.  We divide one day
into 144 10-minute timeslots. We collect the total number of requests and the
total traffic size in each timeslot, and divide them by the number of days
to obtain the averages for each timeslot.  We align the timestamps of all
three traces to their respective time zones: for AliCloud and TencentCloud, we
align the timestamps of the requests to GMT+8, while for MSRC, we align them
to GMT+0.  Recall that the TencentCloud traces end at 1:00~AM on the tenth day
and have a missing hour of requests at 1:00~AM-2:00~AM
(\S\ref{subsec:overview}). The average numbers of requests per day
in the timeslots of 12:00~AM-1:00~AM and 1:00~AM-2:00~AM are obtained by the
total number of requests in the timeslots divided by ten and eight,
respectively; the average amounts of traffic are handled similarly. 

Figures~\ref{fig:per_hour_request} and \ref{fig:per_hour_traffic} show the
distributions of the average numbers of requests and the average amounts of
traffic across different timeslots for all three traces, respectively. 
We see that in both AliCloud and TencentCloud, their average numbers of
requests and sizes of traffic are almost evenly spread
across daytime (6:00~AM to 6:00~PM) and nighttime (6:00~PM to 6:00~AM).  In
AliCloud, the daytime has 52.6\% of the total average number of I/O requests
(Figure~\ref{fig:per_hour_request}(a)) and 52.0\% of total average traffic in
all timeslots (Figure~\ref{fig:per_hour_traffic}(a)), while in TencentCloud,
the daytime has 50.3\% of average I/O requests
(Figure~\ref{fig:per_hour_request}(b)) and 49.6\% of average traffic
(Figure~\ref{fig:per_hour_traffic}(b)). However, in MSRC, the daytime only has
33.7\% of average I/O requests and 23.3\% of average traffic, and the I/O
requests and traffic mainly dominate in nighttime
(Figures~\ref{fig:per_hour_request}(c) and \ref{fig:per_hour_traffic}(c)). 

We also find that from Figures~\ref{fig:per_hour_traffic}(a) and
\ref{fig:per_hour_traffic}(c), AliCloud and MSRC volumes have spikes for reads
at 0:00~AM-0:20~AM and 1:00~AM-2:30~AM, respectively, accounting for 13.1\%
and 18.8\% of all read traffic, respectively; in contrast, TencentCloud does
not have such spikes.  For AliCloud, the reason of the spikes is that 12 out
of the 1,000 volumes only have read requests during 0:00~AM-0:20~AM, and they 
contain significant numbers of large read requests. Each of these 12 volumes
has a total of more than 50.2\,GiB of average read traffic at
0:00~AM-0:20~AM per day.  Their average read request sizes are
larger than 360\,KiB (note that most of the read requests are smaller than
100\,KiB; see Finding~A.2 in \S\ref{subsec:highlevel}).  We suspect that there
exist scheduled scan activities in these 12 volumes at midnight, although we
cannot identify their specific applications (\S\ref{subsec:overview}).  If we
exclude these 12 volumes, the overall intensities of read traffic for AliCloud
at midnight will be comparable to other time intervals. 
For MSRC, the reason of the spikes near midnight is that a volume called {\em
src1\_1} has an extremely large amount of read traffic at 1:00~AM-2:30~AM,
accounting for 56.8-82.9\% of the read traffic of all volumes in MSRC in the
corresponding 10-minute time intervals. Note that the average read request
sizes of {\em src1\_1} are smaller than 43.1\,KiB during the spike period, as
opposed to the large requests in AliCloud. Read spikes near
midnight are common in production; for example, in prior work
\cite{kavalanekar08}, massive read spikes are observed at about 3:30~AM in a
production server due to the scheduled replication tasks in early mornings.

\subsection{Spatial Patterns}
\label{subsec:spatial}

We study the spatial characteristics of volumes in AliCloud, TencentCloud, and
MSRC through the following metrics.  First, we study the randomness of I/O
requests by examining the offset differences of recent requests \cite{ahmad07,
tarihi15}, as random I/Os can compromise the performance and endurance of
flash-based storage \cite{min12}.  Second, we examine the aggregations of reads
and writes in working sets, so as to provide guidelines for resource
allocation in caching \cite{soundararajan10, li16, li19}.  Finally, we examine
the patterns of update coverage (i.e., the percentage of WSS for updates),
which is important for optimizing update performance in storage cluster
management \cite{chan14}.

\paragraph{Finding~B.9:} {\em Random I/Os are common in AliCloud,
TencentCloud, and MSRC. The volumes in AliCloud and TencentCloud see higher
percentages of random I/Os than those in MSRC.}

We study the randomness of I/O requests by examining the spatial relationships
among adjacent requests.  To quantify the randomness of a request, we measure
the minimum distance between the current offset of the request and the offsets
of the previous 32 requests \cite{ahmad07,tarihi15}.  If the minimum distance
exceeds a threshold (e.g., 128\,KiB, which is the read-ahead length of the
surveyed drives in \cite{tarihi15}), we regard the request as a random
request. We measure the {\em randomness ratio} of a volume, defined as the
percentage of random requests over all requests.

\begin{figure}[!t] 
\centering
\begin{tabular}{@{\ }cc}
\includegraphics[width=2.3in]{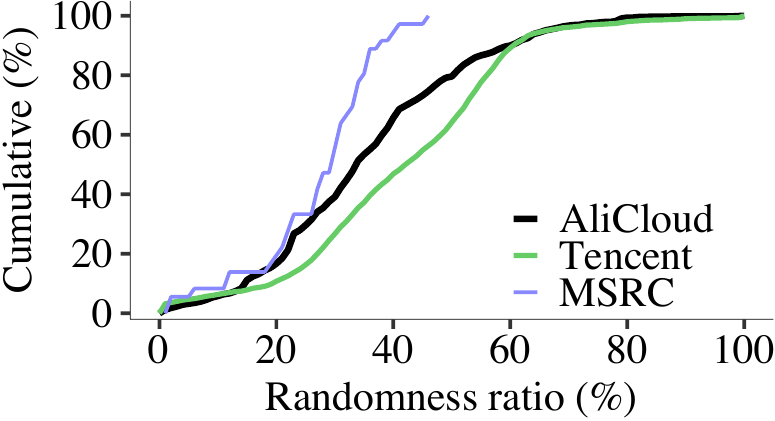} &
\includegraphics[width=2.3in]{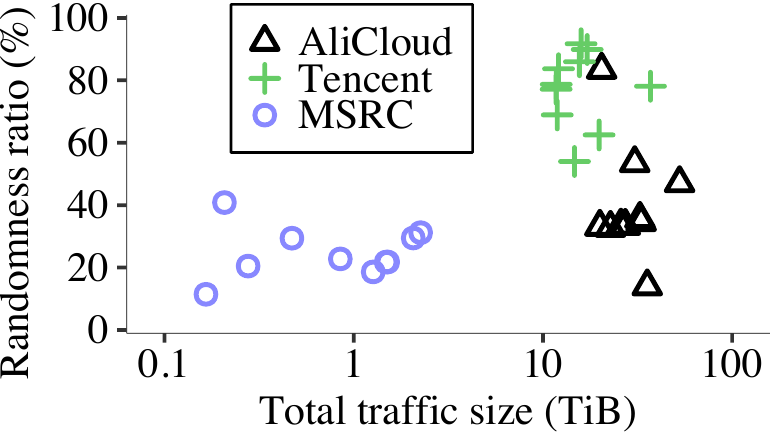} \\
{\small (a) Randomness ratio distribution}  & 
{\small (b) Randomness ratio vs. traffic}
\end{tabular}
\vspace{-3pt}
\caption{Finding~B.9: Cumulative distributions of randomness ratios of volumes
(figure~(a)) and the relationship between the randomness ratios and total
traffic sizes in top-10 traffic-intensive volumes.}
\label{fig:randomness}
\end{figure}

Figure~\ref{fig:randomness}(a) shows the cumulative distributions of
randomness ratios of volumes in AliCloud, TencentCloud, and MSRC. We find that
random I/Os are common in all three traces.  Half of the volumes have at least
33.5\%, 42.1\%, and 29.4\% of random I/Os in AliCloud, TencentCloud, and MSRC,
respectively. Also, AliCloud and TencentCloud in general show higher
randomness ratios than MSRC. In particular, all volumes in MSRC have less than
46\% of random requests, while 20.4\% and 35.8\% of volumes in AliCloud and
TencentCloud have more than 50\% of random requests, respectively.  The
existence of randomness may come from file system workloads as witnessed in
\cite{he17}.

We further examine the randomness ratios of the top-10 volumes that have the
most I/O traffic in each trace.  Figure~\ref{fig:randomness}(b) shows the
relationships between the randomness ratios and the total I/O traffic sizes of
the top-10 volumes. We see that the volumes with large amounts of I/O traffic
have high randomness ratios in general. The randomness ratios of the top-10
volumes in AliCloud, TencentCloud, and MSRC are 14.0-83.4\%, 54.0-91.7\%, and
11.4-40.8\%, respectively, and their I/O traffic sizes are 20.0-52.8\,TiB,
11.7-36.8\,TiB, and 0.17-2.26\,TiB, respectively. We also examine
the Spearman correlation coefficients \cite{spearman87} between the I/O
traffic sizes and the randomness ratios in the top-10 volumes of all three
traces.  We find that the coefficients are 0.079, 0.164, and 0.333 in
AliCloud, TencentCloud, and MSRC, respectively, implying that there exist
positive correlations between the two metrics in the top-10 volumes. The
results indicate that random I/Os are common in traffic-intensive volumes.

Combining with the observation that small-size I/O requests dominate in all
three traces (\S\ref{subsec:highlevel}), we see that random and small I/Os are
common in all three traces, especially in AliCloud and TencentCloud.  Such
access patterns can compromise the performance and endurance of flash-based
storage \cite{min12, he17}.

\begin{figure}[!t] 
\centering
\begin{tabular}{@{\ }ccc}
\includegraphics[width=2in]{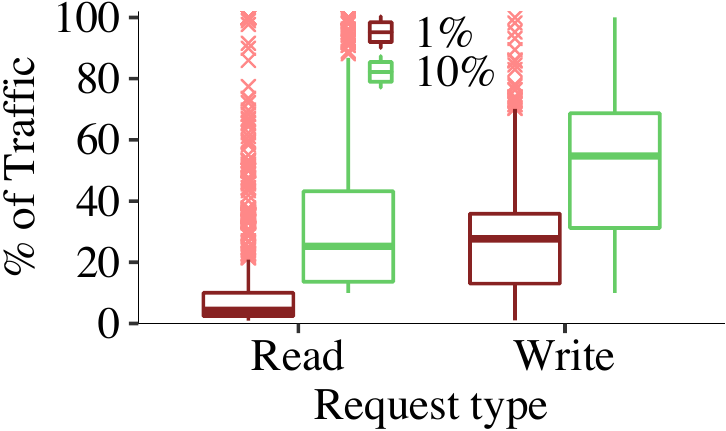} &
\includegraphics[width=2in]{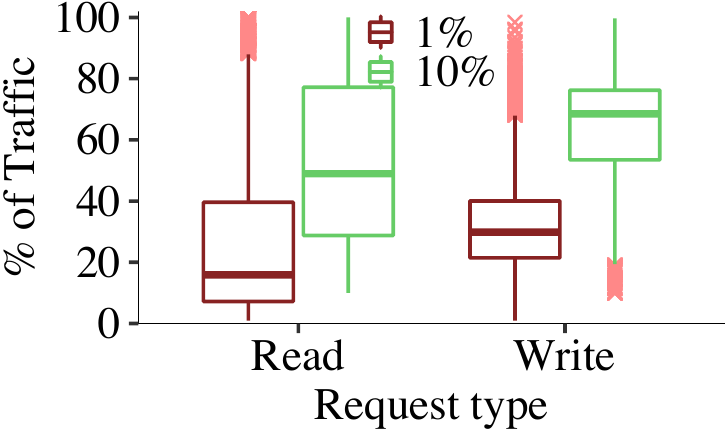} &
\includegraphics[width=2in]{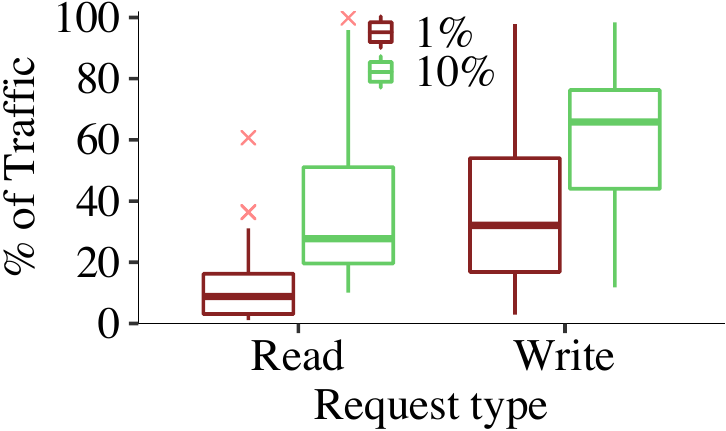} \\
{\small (a) AliCloud}  & 
{\small (b) TencentCloud}  & 
{\small (c) MSRC} 
\end{tabular}
\vspace{-3pt}
\caption{Finding~B.10: Boxplots of percentages of traffic sizes for the top-1\%
and top-10\% read and write blocks across all volumes.}
\label{fig:skewness}
\end{figure}

\paragraph{Finding~B.10: } {\em Reads and writes are aggregated in small working
sets in non-negligible fractions of volumes in AliCloud, TencentCloud, and
MSRC, while TencentCloud has the highest aggregation of reads among all three
traces.  Writes are more aggregated than reads.}

We study how reads and writes are aggregated in the working sets of each volume.
Specifically, in the read (or write) working sets, we focus on the top-1\% and
top-10\% of unique blocks that receive the most read (or write)
traffic.  We examine the percentage of the read (or write) traffic size of
each such block over the total read (or write) traffic size; a higher
percentage implies that the I/O traffic is more aggregated in a block. 

Figure~\ref{fig:skewness} shows the boxplots of percentages of traffic sizes
for the top-1\% and top-10\% blocks across all volumes in AliCloud,
TencentCloud, and MSRC. We first focus on read traffic. We see that
read traffic is aggregated in the top-1\% and top-10\% blocks in
non-negligible fractions of volumes. In AliCloud, 75\% of volumes have at
least 2.5\% and 13.6\% of read traffic in the top-1\% and top-10\% read
blocks, respectively (Figure~\ref{fig:skewness}(a)).  In TencentCloud, the
corresponding percentages are 7.2\% and 28.8\%, respectively
(Figure~\ref{fig:skewness}(b)), and in MSRC, the corresponding percentages are
3.1\% and 19.6\%, respectively (Figure~\ref{fig:skewness}(c)).  In particular,
the aggregation for reads is the highest in TencentCloud among all three
traces.

In AliCloud, the boxplots indicate 147 volumes as outliers in the top-1\% read
blocks (Figure~\ref{fig:skewness}(a)).  Such outlier volumes have more than
21.3\% of read traffic in their top-1\% read blocks.  It implies that a small
read cache can absorb a substantial amount of read traffic for such volumes. 

Compared with reads, writes are more aggregated.  In AliCloud, the 25th
percentiles of read traffic in the top-1\% and top-10\% read blocks are
2.5\% and 13.6\%, respectively, while the 25th percentiles of write traffic in
the top-1\% and top-10\% written blocks increase to 13.0\% and 31.2\%,
respectively (Figure~\ref{fig:skewness}(a)). Similar observations hold in
TencentCloud and MSRC. For example, in TencentCloud, the 25th percentiles of
read and write traffic in the top-10\% read and written blocks are 28.8\% and
53.5\%, respectively (Figure~\ref{fig:skewness}(b)), and in MSRC, the
corresponding 25th percentiles are 19.6\% and 44.0\%, respectively
(Figure~\ref{fig:skewness}(c)).  Note that the spatial clustering of writes
is common in desktop applications and is related to files such as mail boxes,
search indexes, and file system metadata \cite{soundararajan10}.

\paragraph{Finding~B.11:} {\em Reads and writes tend to be aggregated in
read-mostly and write-mostly blocks, respectively, in AliCloud and
TencentCloud.}

We further classify the blocks into different types as in \cite{li16} and
examine the aggregation of reads and writes.  Specifically, we classify a block
as {\em read-mostly} (or {\em write-mostly}) if its read (or write) traffic
occupies more than 95\% of its total I/O traffic.  We examine the percentage
of all read (or write) traffic in the whole trace duration that goes to
read-mostly (or write-mostly) blocks. 

Table~\ref{tab:rw_heavy} shows the overall percentages of all read and write
traffic that goes to read-mostly and write-mostly blocks in AliCloud,
TencentCloud, and MSRC.  In AliCloud and TencentCloud, the majority of read
traffic (59.1\% and 78.5\%, respectively) and write traffic (80.7\% and 90.8\%,
respectively) goes to read-mostly blocks and write-mostly blocks, respectively.
In MSRC, 72.1\% of read traffic goes to read-mostly blocks; however, only
32.4\% of write traffic goes to write-mostly blocks. Overall, both AliCloud
and TencentCloud show prominent aggregations of reads and writes in
read-mostly and write-mostly blocks, respectively, but it is not the case in
MSRC.  Note that the limited aggregation of writes in write-mostly blocks in
MSRC is inconsistent with the prior finding in \cite{li16}. The reason is that
the study in \cite{li16} considers only 12 out of 36 volumes in MSRC, while we
consider all 36 volumes.

\begin{table}[t]
\small
\centering
\renewcommand{\arraystretch}{1.1}
\begin{tabular}{c|c|c|c}
\hline
Traces       & AliCloud           & TencentCloud & MSRC \\ 
\hline
\hline
Reads to read-mostly blocks (\%) & 59.2       & 78.5   & 75.9           \\ 
\hline
Writes to write-mostly blocks (\%) & 80.7     & 90.8   & 33.5           \\ 
\hline
\end{tabular}
\caption{Finding~B.11: Percentages of all read and write traffic going to
read-mostly and write-mostly blocks, respectively.}
\label{tab:rw_heavy}
\end{table}

\begin{figure}[!t] 
\centering
\begin{tabular}{@{\ }ccc}
\includegraphics[width=2in]{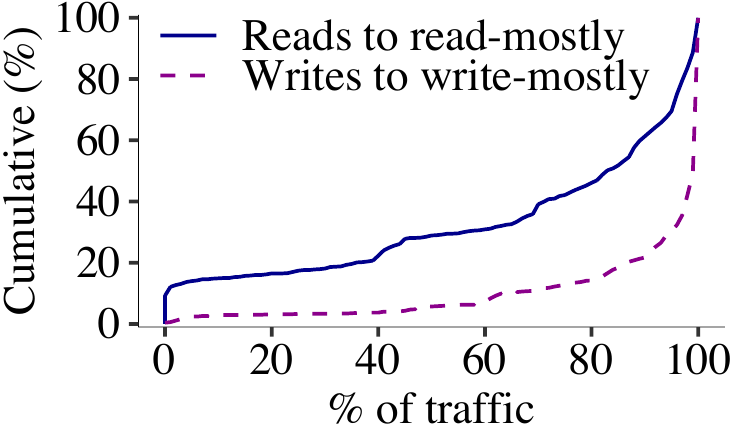} &
\includegraphics[width=2in]{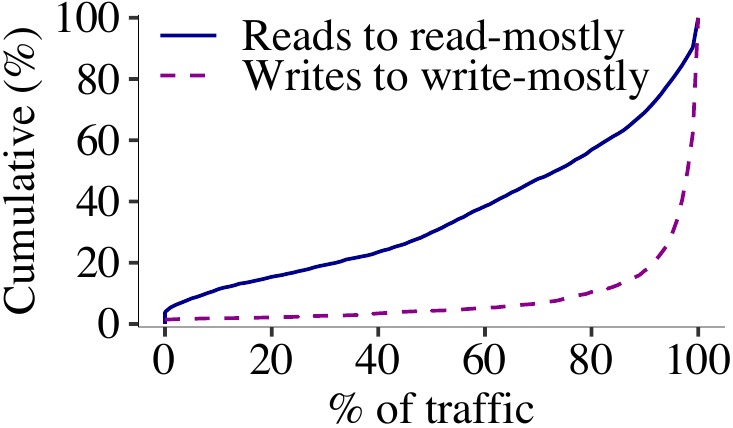} &
\includegraphics[width=2in]{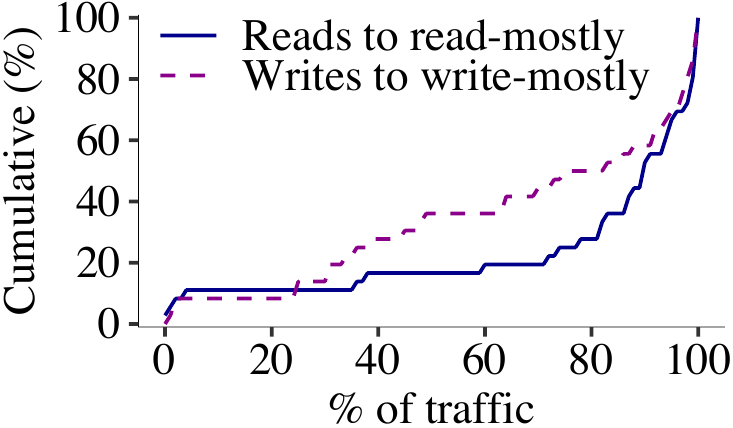} \\
{\small (a) AliCloud}  & 
{\small (b) TencentCloud}  & 
{\small (c) MSRC}
\end{tabular}
\vspace{-3pt}
\caption{Finding~B.11: Cumulative distributions of all percentages of read and
write traffic going to read-mostly and write-mostly blocks, respectively,
across all volumes.} 
\label{fig:rw_heavy}
\end{figure}

Figure~\ref{fig:rw_heavy} shows the cumulative distributions of percentages of
read and write traffic that goes to read-mostly and write-mostly blocks,
respectively, across all volumes in AliCloud, TencentCloud, and
MSRC. Most of the volumes in all three traces have high percentages of all
read and write traffic aggregated in read-mostly and write-mostly blocks,
respectively.  In AliCloud, half of the volumes have more than 82.6\% of reads
going to read-mostly blocks and more than 99.2\% writes going to write-mostly
blocks (Figure~\ref{fig:rw_heavy}(a)); in TencentCloud, the corresponding
percentages are 73.4\% and 98.0\%, respectively (Figure~\ref{fig:rw_heavy}(b)); 
in MSRC, the corresponding percentages are 89.4\% and 78.3\%, respectively
(Figure~\ref{fig:rw_heavy}(c)).

\paragraph{Finding~B.12:} {\em AliCloud and TencentCloud generally have higher
update coverages and higher percentages of update traffic than MSRC.  The
update coverage also varies across volumes.}

Recall that Table~\ref{tab:basic} (Section~\ref{subsec:highlevel}) shows the
overall WSSs (working set sizes) for reads, writes, and updates.  We now
examine the spatial characteristics of updates.  We focus on the update working
set, which covers the blocks that are written more than once.  We measure the
{\em update coverage} of a volume, defined as the ratio between the update WSS
and the total WSS of the volume \cite{chan14}. In addition, we measure the
percentages of update traffic over the total amount of traffic across all
volumes. 

Table~\ref{tab:update} shows the averages, medians, and 90th
percentiles of update coverages of all volumes in all three traces. In general,
AliCloud and TencentCloud have higher update coverages than MSRC. In AliCloud
and TencentCloud, half of the volumes have update coverages of more than 61.2\%
and 56.7\%, respectively, while in MSRC, the corresponding percentage is 9.4\%
only. In addition, Table~\ref{tab:update_traffic} shows that AliCloud and
TencentCloud have higher percentages of update traffic than MSRC in terms
of the averages, medians, and 90th percentiles.  This suggests that AliCloud
and TencentCloud are more update-intensive than MSRC. 

\begin{figure}[!t] 
\centering
\begin{tabular}{@{\ }cc}
\multicolumn{2}{c}{\includegraphics[width=2in]{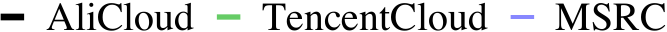}} \\
\includegraphics[width=2.3in]{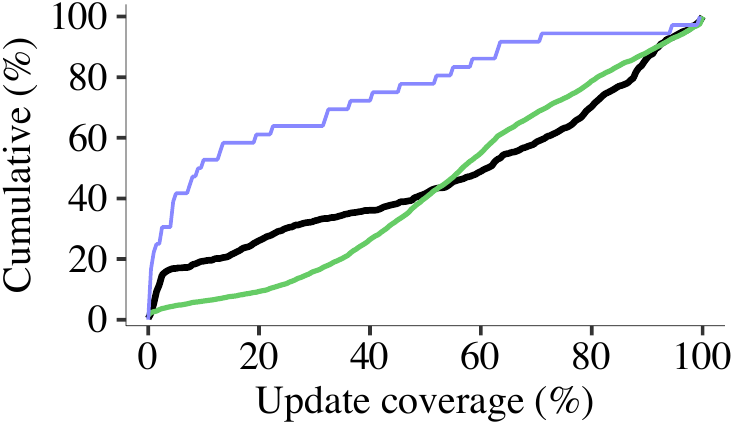} &
\includegraphics[width=2.3in]{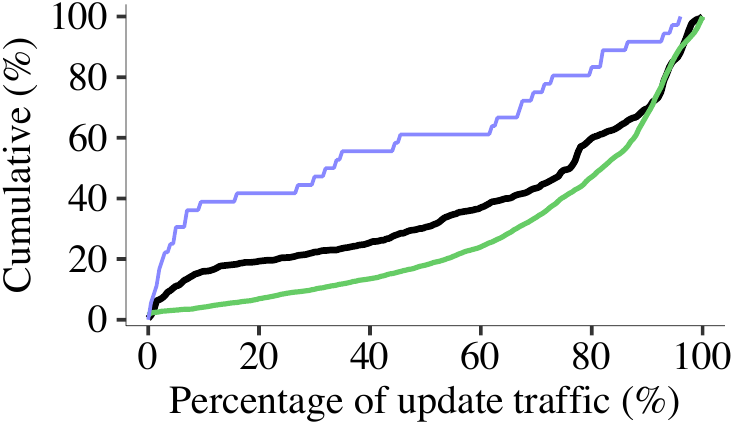} \\ 
{\small (a) Update coverages}  & 
{\small (b) Percentages of update traffic}
\end{tabular}
\vspace{-3pt}
\caption{Finding~B.12: Cumulative distributions of update coverages
and percentages of update traffic across all volumes. }
\label{fig:update}
\end{figure}

Figure~\ref{fig:update}(a) shows the cumulative distributions of update
coverages across all volumes in AliCloud, TencentCloud,
and MSRC. In AliCloud and TencentCloud, 45.2\% and 37.2\% of volumes have
update coverages larger than 65\%, respectively, and their update coverages
are more diverse than MSRC.  On the other hand, in MSRC, 33 out of
36 volumes have update coverages below 65\%. 
Figure~\ref{fig:update}(b) shows the cumulative distributions of percentages
of update traffic. AliCloud and TencentCloud show higher percentages of update
traffic than MSRC, and TencentCloud generally has a higher percentage than
AliCloud. 

\begin{table}[t]
  \begin{minipage}{3in} \small
    \centering
    \renewcommand{\arraystretch}{1.1}
    \setlength\tabcolsep{3pt} 
    \begin{tabular}{c|c|c|c}
    \hline
    Traces          &  AliCloud     & TencentCloud &  MSRC \\ 
    \hline
    \hline
    Mean (\%)  &  52.7    & 56.2 & 24.1 \\ 
    \hline
    Median (\%)   &  61.2    & 56.7 & 9.4  \\ 
    \hline
    90th PCTL (\%) &  92.1   & 91.9 & 63.0  \\ 
    \hline
    \end{tabular}
    \caption{Finding~B.12: Means, medians and 90th percentiles of
    update coverages of all volumes.}
    \label{tab:update}
  \end{minipage}
  \hspace{0.2in}
  \begin{minipage}{3in} \small
    \centering
    \renewcommand{\arraystretch}{1.1}
    \setlength\tabcolsep{3pt} 
    \begin{tabular}{c|c|c|c}
    \hline
    Traces          &  AliCloud     & TencentCloud &  MSRC \\ 
    \hline
    \hline
    Mean (\%)  &  62.5    & 72.5 & 38.7 \\ 
    \hline
    Median (\%)   &  76.0    & 81.9 & 32.7 \\ 
    \hline
    90th PCTL (\%) &  96.4    & 96.1 & 84.2 \\ 
    \hline
    \end{tabular}
    \caption{Finding~B.12: Means, medians, and 90th percentiles
    of the percentages of update traffic of all volumes.}
    \label{tab:update_traffic}
  \end{minipage}
\end{table}

\subsection{Temporal Patterns}
\label{subsec:temporal}

We study the temporal characteristics of volumes in AliCloud,
TencentCloud, and MSRC by examining the temporal relationships of adjacent I/O
requests. We first examine the time elapsed between adjacent requests to the
same block with respect to different combinations of read and write requests
for workload-aware caching designs \cite{soundararajan10}.  We also study the
update interval (i.e., the time interval between two consecutive writes to the
same block), which facilitates flash-based storage management \cite{liu12,
cai15}.  Finally, we study the miss ratios under least recently used (LRU)
caching, which reflects the temporal aggregation of traffic for caching
efficiency \cite{wires14, waldspurger15}.  

Recall that the TencentCloud traces have a missing hour of requests at
1:00~AM-2:00~AM on the eighth day (\S\ref{subsec:overview}).  Thus, in our
following analysis for TencentCloud, we discard the adjacent requests that
span across the missing hour. 

\paragraph{Finding~B.13:} {\em  The read-after-write (RAW) times in AliCloud,
TencentCloud, and MSRC are generally larger than the write-after-write (WAW)
times.  Also, TencentCloud generally has smaller RAW times than AliCloud and
MSRC, while AliCloud generally has larger WAW times than TencentCloud and
MSRC.  Furthermore, AliCloud and TencentCloud have significantly more WAW
requests than RAW requests.}

We first examine two types of adjacent requests \cite{soundararajan10}: (i) a
{\em read-after-write (RAW)} request, which refers to the read following
immediately the write to the same block; and (ii) a {\em write-after-write
(WAW)} request, which refers to the write following immediately the write to
the same block.  We measure the time of a RAW (resp.  WAW) request as the
elapsed time between the adjacent read and write (resp. the two adjacent
writes) to the same block. 

\begin{figure}[!t] 
\centering
\begin{tabular}{@{\ }ccc}
\includegraphics[width=2in]{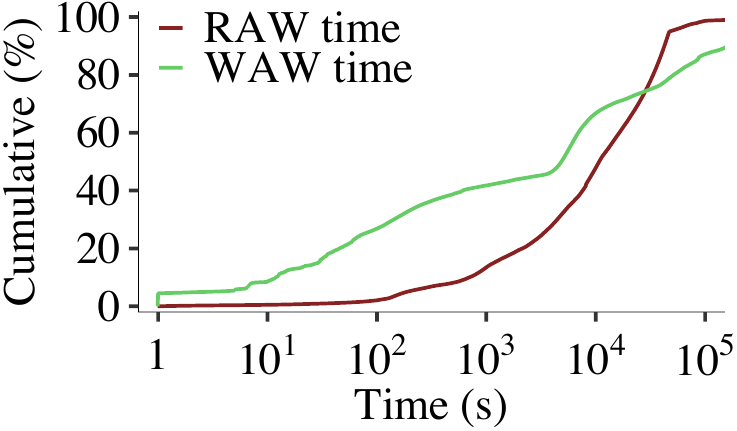} &
\includegraphics[width=2in]{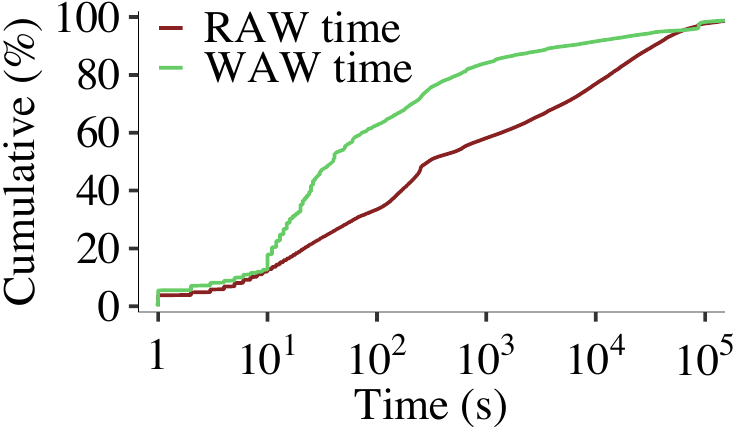} &
\includegraphics[width=2in]{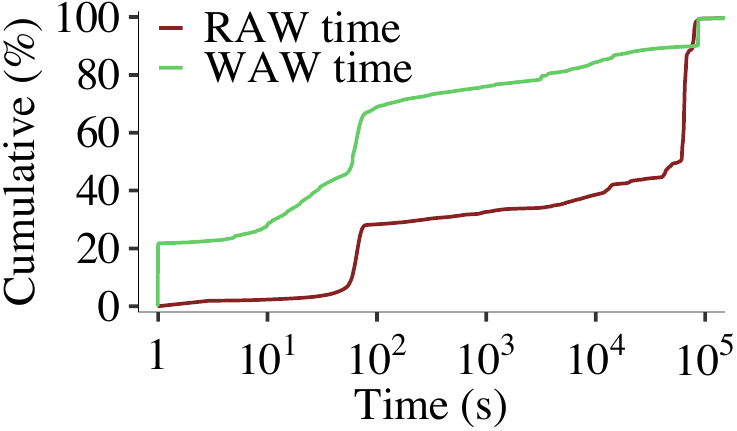} \\
{\small (a) AliCloud}  & 
{\small (b) TencentCloud}  & 
{\small (c) MSRC} 
\vspace{3pt}\\
\includegraphics[width=2in]{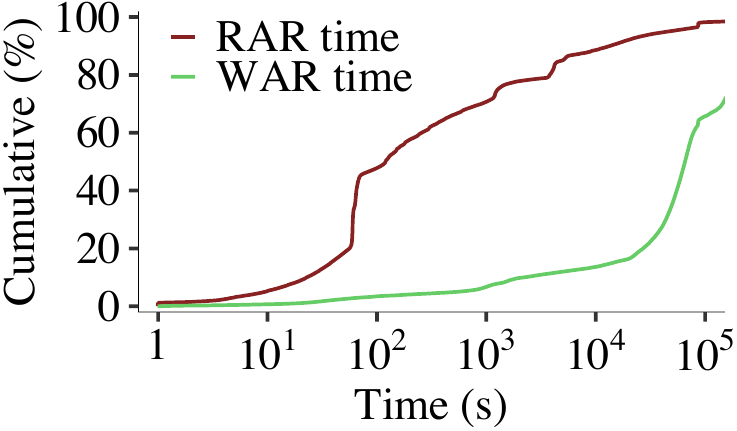} &
\includegraphics[width=2in]{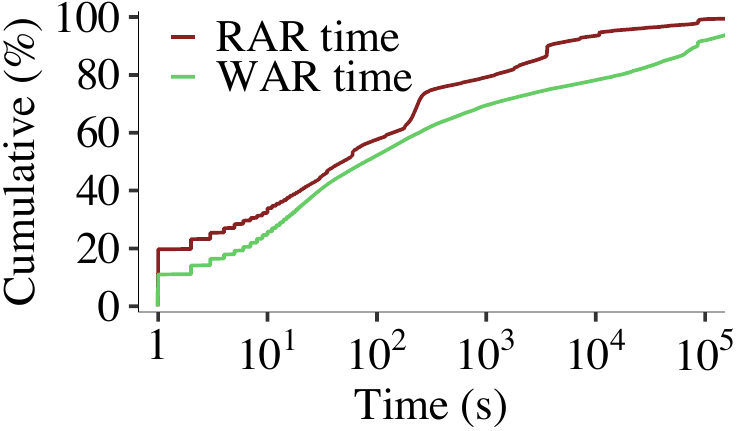} &
\includegraphics[width=2in]{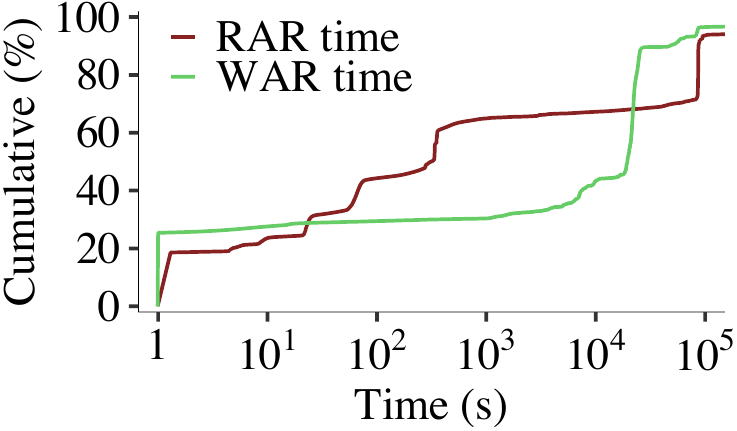} \\
{\small (d) AliCloud}  & 
{\small (e) TencentCloud}  & 
{\small (f) MSRC}
\end{tabular}
\vspace{-3pt}
\caption{Findings~B.13-B.14: Cumulative distributions of RAW, WAW,
RAR, and WAR times across all RAW, WAW, RAR, and WAR requests, respectively.}
\label{fig:rwarw}
\end{figure}

Figures~\ref{fig:rwarw}(a)-\ref{fig:rwarw}(c) show the cumulative
distributions of RAW and WAW times across all RAW and WAW requests,
respectively, in all three traces.  All three traces generally have larger RAW
times than WAW times.  Specifically, the 50th percentiles of the RAW time in
AliCloud, TencentCloud, and MSRC are 3.0~hours, 4.9~minutes, and 16.1~hours,
respectively, while the 50th percentiles of the WAW time are only 1.3~hours,
0.7~minutes, and 1.0~minute, respectively.  Such findings are consistent with
those in prior studies \cite{hsu03,soundararajan10}.  As for the possible
reasons, the smaller WAW times are likely to appear in desktop workloads
\cite{soundararajan10}, while the larger RAW times are possibly related to the
large OS-level buffer caches \cite{soundararajan10}. 

Also, all three traces have different percentages of large and small RAW and
WAW times.  To aid our analysis, we treat the times smaller than 1.0~minute
and larger than 15.0~minutes as small and large, respectively, as also used in
\cite{soundararajan10}. For RAW times, TencentCloud has the largest percentage
of RAW times smaller than 1.0~minute among all three traces, while most of the
RAW times in AliCloud and MSRC are larger than 15.0~minutes.  Specifically, 
1.4\%, 29.8\%, and 10.9\% of the RAW times are smaller than 1.0~minute in
AliCloud, TencentCloud, and MSRC, respectively, while 87.8\%, 42.5\%, and
67.8\% of the RAW times are larger than 15.0~minutes in the three traces,
respectively.  On the other hand, most of the WAW times are smaller than
1.0~minute in TencentCloud and MSRC, while most of the WAW times are larger
than 15.0~minutes in AliCloud.  Specifically, 22.7\%, 57.7\%, and 51.5\% of
WAW times are smaller than 1.0~minute in AliCloud, TencentCloud, and MSRC,
respectively, while 58.5\%, 16.4\%, and 24.3\% of WAW times are larger than
15.0~minutes in the three traces, respectively. 

Table~\ref{tab:rwarw} shows the numbers of RAW and WAW requests in AliCloud,
TencentCloud, and MSRC. We observe a large difference in the numbers of RAW
and WAW requests in both AliCloud and TencentCloud, but a small difference in
MSRC. Specifically, in AliCloud, the numbers of RAW and WAW requests are
12.4~billion and 103.7~billion, respectively; the number of WAW requests is
8.3$\times$ that of RAW requests. In TencentCloud, the numbers of RAW and WAW
requests are 8.8~billion and 204.9~billion, respectively, and the number of
WAW requests is even 23.3$\times$ that of RAW requests. In MSRC, those
numbers are 297.2~million and 289.2~million, respectively, and are close to
each other. We suspect that the larger number of WAW requests in
AliCloud and TencentCloud may be related to the metadata and log files in the
workloads; for example, the fraction of WAW requests drops significantly after
the metadata and log files are excluded \cite{hsu03}. 

\paragraph{Finding~B.14: } {\em TencentCloud has the highest fractions of small
RAR and WAR times and the smallest fractions of large RAR and WAR times in all
three traces.  There exist extremely small RAR and WAR times in MSRC.  In all
three traces, the WAR time is much larger than the RAR time, and there are
much more RAR requests than WAR requests.}

We further examine two types of adjacent requests: (i) a {\em read-after-read
(RAR)} request, which refers to the read following immediately the read to the
same block; and (ii) a {\em write-after-read (WAR)} request, which refers to
the write following immediately the read to the same block.  

Figures~\ref{fig:rwarw}(d)-\ref{fig:rwarw}(f) show the cumulative
distributions of RAR and WAR times across all RAR and WAR requests,
respectively, in all three traces.  We again treat the times smaller than
1.0~minute and larger than 15.0~minutes as small and large, respectively, as
above.  TencentCloud has the highest fractions of RAR and WAR times smaller
than 1.0~minute, and the lowest fractions of RAR and WAR times larger than
15.0~minutes.  Specifically, 28.5\%, 53.4\%, and 35.6\% of the RAR times are
smaller than 1.0~minute in AliCloud, TencentCloud, and MSRC, respectively,
while 30.0\%, 21.2\%, and 35.2\% of the RAR times are larger than
15.0~minutes, respectively.  On the other hand, 2.8\%, 47.6\%, and 29.2\% of
the WAR times are smaller than 1~minute in AliCloud, TencentCloud, and MSRC, 
respectively, while 93.8\%, 31.1\%, and 69.7\% of the WAR times are larger
than 15~minutes, respectively.  
In particular, in MSRC, there exist non-negligible
fractions of extremely small RAR and WAR times (18.5\% and 25.4\%,
respectively) that are smaller than 1~second, which is not the case for
AliCloud and TencentCloud. 

Overall, in all three traces, the WAR time is much larger than the RAR
time. In AliCloud, the 50th percentiles of RAR and WAR times are 2.0~minutes
and 18.2~hours, respectively, and 21.0\% and 88.8\% of RAR and WAR times are
larger than 1~hour, respectively (Figure~\ref{fig:rwarw}(d)). In TencentCloud,
the 50th percentiles of RAR and WAR time are 49~seconds and 78~seconds,
respectively, and 11.1\% and 25.2\% of RAR and WAR times are larger than
1~hour, respectively (Figure~\ref{fig:rwarw}(e)).  In MSRC, the 50th
percentiles of RAR and WAR times are 5.2~minutes and 5.4~hours, respectively,
and 33.6\% and 66.7\% of RAR and WAR times are larger than 1~hour,
respectively (Figure~\ref{fig:rwarw}(f)). The results indicate that a block
being read is likely read again soon.  

We also examine the numbers of RAR and WAR requests in AliCloud, TencentCloud,
and MSRC, as shown in Table~\ref{tab:rwarw}. In AliCloud, TencentCloud, and
MSRC, the numbers of RAR requests are 2.54$\times$, 8.07$\times$, and
4.19$\times$ those of WAR requests, respectively.

\begin{table}[t]
\small
\centering
\renewcommand{\arraystretch}{1.1}
\begin{tabular}{c|c|c|c|c}
\hline
Traces &  RAW (M)  & WAW (M) & RAR (M) & WAR (M)  \\ 
\hline
\hline
AliCloud     & 12,432.7 & 103,708.4 & 29,845.0 & 11,760.6  \\ 
\hline
TencentCloud & 8,796.0  & 204,856.2 & 63,990.4 & 7,930.3  \\ 
\hline
MSRC         & 297.2    & 289.2     & 1,382.4  & 330.0  \\ 
\hline
\end{tabular}
\caption{Findings~B.13-B.14: Numbers of RAW, WAW, RAR, and WAR requests (in
millions).}
\label{tab:rwarw}
\end{table}

\paragraph{Finding~B.15:} {\em Written blocks have varying update intervals.}

We measure the {\em update interval} of a block, defined as the elapsed time
between two consecutive writes to the same block.  Note that the update
interval differs from the WAW time, as the former allows reads between two
writes.  Each block may be written more than once, so it may be associated with
multiple update intervals (e.g., a block that is written $M$ times has $M-1$
update intervals).  The update interval of a block describes the lifetime of
the block data.  
   
Table~\ref{tab:update_interval} shows different percentiles of update
intervals across all volumes in all three traces. In AliCloud, the update
intervals generally have long durations, while in TencentCloud and MSRC, the
update intervals are generally small, especially in TencentCloud. In AliCloud,
50\% of update intervals are larger than 95.2~minutes (1.6~hours), and the
90th percentile is 3,017.4~minutes (50.3~hours).  In TencentCloud, the 25th,
50th, and 75th percentiles are only several seconds or minutes (0.23~minutes,
0.67~minutes, and 5.4~minutes, respectively), implying that the majority of
updated blocks have extremely high update frequencies. However, some updated
blocks still have high update intervals, as the 90th and 95th
percentiles are 120.0~minutes (2.0~hours) and 973.1~minutes (16.2~hours),
respectively.  In MSRC, the update intervals have a bimodal pattern, in which
50\% of update intervals are smaller than 1.25~minutes, while 25\% of update
intervals are larger than 1,438.9~minutes (24.0~hours). The reason of such a
bimodal pattern in MSRC is that a volume is responsible for source control
(i.e., {\em src1\_0}) and updates data blocks daily. If we exclude the daily
updates, most of the written blocks in MSRC have very short update intervals.

\begin{figure}[!t] 
\centering
\begin{tabular}{@{\ }ccc}
\includegraphics[width=2in]{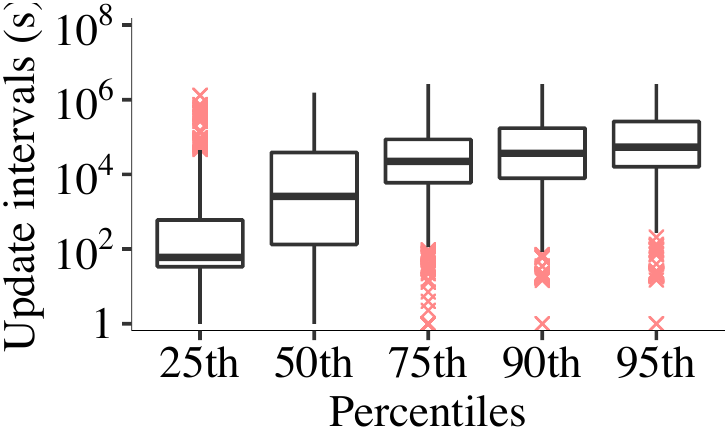} &
\includegraphics[width=2in]{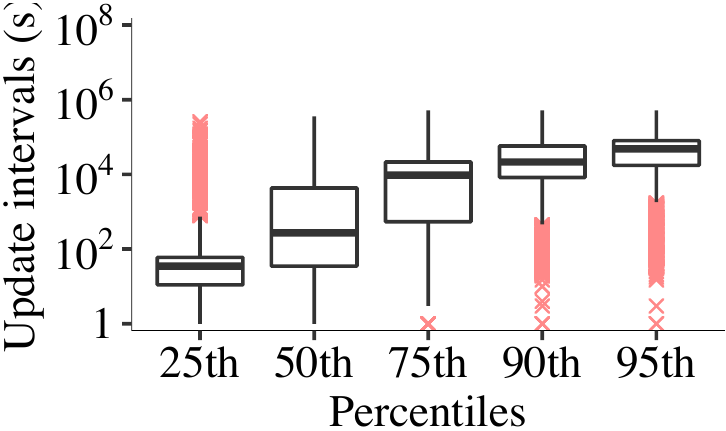} &
\includegraphics[width=2in]{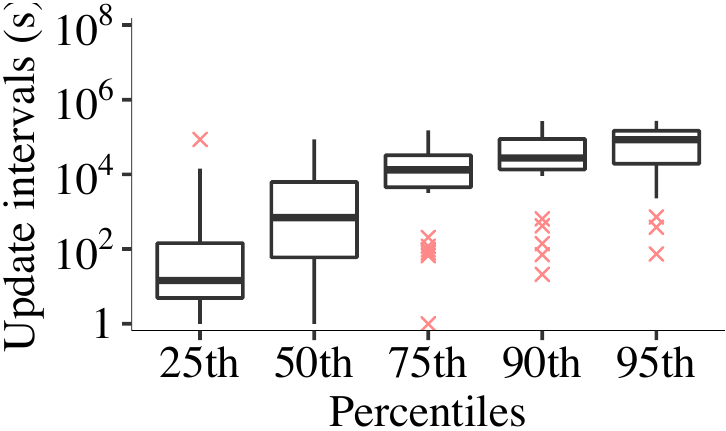} \\
{\small (a) AliCloud}  & 
{\small (b) TencentCloud}  & 
{\small (c) MSRC}
\end{tabular}
\vspace{-3pt}
\caption{Finding~B.15: Boxplots of percentiles of update intervals across all
volumes.}
\label{fig:update_interval}
\end{figure}

Figure~\ref{fig:update_interval} shows the boxplots of update intervals of
different groups of percentiles across all volumes in AliCloud,
TencentCloud, and MSRC.  We see that the distributions of update intervals
have high variations across volumes in all three traces.
For example, in AliCloud, the 50th percentiles of update intervals of all
volumes range from 1~second to 17.8~days (Figure~\ref{fig:update_interval}(a));
in TencentCloud, the 50th percentiles vary between 1~second and
4.14~days (Figure~\ref{fig:update_interval}(b)); in MSRC, the 50th percentiles
of update intervals of all volumes range from 1~second to 24~hours.

Many volumes have non-negligible proportions of short update intervals in their
update requests.  To further examine the distributions of update intervals in
individual volumes, we divide the update intervals into four groups: (i) less
than 5~minutes, (ii) 5-30~minutes, (iii) 30-240 minutes, and (iv) more than
240~minutes. We calculate the proportions for the four groups of update
intervals for each volume, and represent the proportions across all volumes by
boxplots. 

\begin{figure}[!t] 
\centering
\begin{tabular}{@{\ }ccc}
\includegraphics[width=2in]{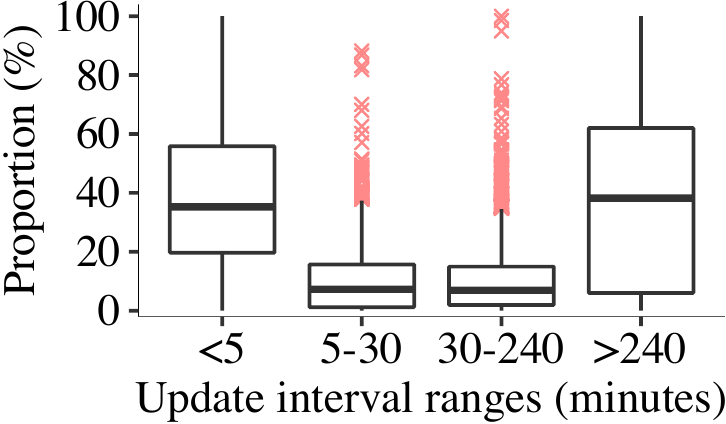} &
\includegraphics[width=2in]{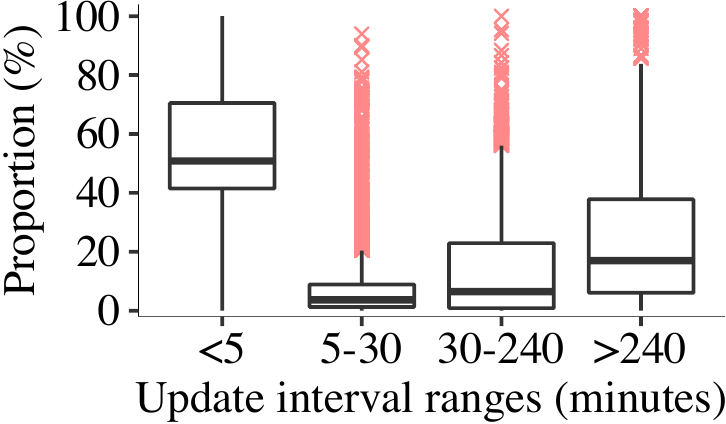} &
\includegraphics[width=2in]{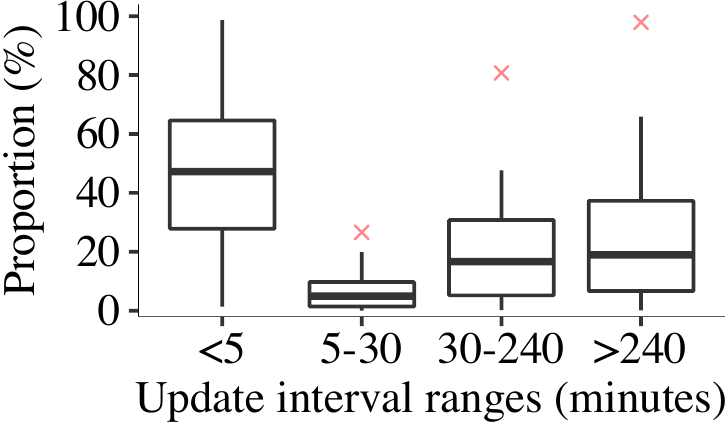} \\
{\small (a) AliCloud}  & 
{\small (b) TencentCloud}  & 
{\small (c) MSRC}
\end{tabular}
\vspace{-3pt}
\caption{Finding~B.15: Boxplots of proportions for the four groups of update
intervals across all volumes.} 
\label{fig:update_interval_levels}
\end{figure}

Figure~\ref{fig:update_interval_levels} shows the boxplots of proportions for
the four groups of update intervals across all volumes in all 
three traces.  All three traces have large proportions
of either very small or very large update intervals. In AliCloud, half of the
volumes have more than 35.2\% and 38.2\% of update intervals in less than
5~minutes and in more than 240~minutes, respectively
(Figure~\ref{fig:update_interval_levels}(a)). In TencentCloud, the
corresponding percentages are 50.8\% and 17.0\%, respectively
(Figure~\ref{fig:update_interval_levels}(b)), while in MSRC, the corresponding
percentages are  47.2\% and 18.9\%, respectively 
(Figure~\ref{fig:update_interval_levels}(c)). Thus, a substantial amount of
data is either updated frequently or not updated for long. 

\begin{table}[t]
\small
\centering
\renewcommand{\arraystretch}{1.1}
\begin{tabular}{c|c|c|c|c|c}
\hline
Percentiles (minutes)  &  25th    &  50th    & 75th  & 90th  & 95th  \\ 
\hline
\hline
AliCloud         &  1.86   &  95.2   & 926.3   &  3,017.4  & 7,200.5  \\
\hline
TencentCloud     &  0.23   &  0.67   & 5.4    &  120.0   &  973.1 \\
\hline
MSRC             &  0.73   &  1.25   & 1,438.9 &  1,440.5  & 1,444.1 \\
\hline
\end{tabular}
\caption{Finding~B.15: Overall percentiles of update intervals across all
volumes.}
\label{tab:update_interval}
\end{table}

\paragraph{Finding~B.16.} {\em Many volumes in TencentCloud have low
miss ratios even under a small cache size, while there are fewer such volumes
with low miss ratios in AliCloud and MSRC.  Also, when the cache size
increases, AliCloud and TencentCloud show the highest absolute reductions in
read and write miss ratios, respectively, in all three traces.}

Finally, we study the impact of caching with respect to the temporal patterns
of the volumes.  For each volume, we simulate a fixed-size cache for both reads
and writes using the LRU policy, and evaluate the corresponding cache miss
ratios for reads and writes.  Here, we select 1\% and 10\% of the WSS of a
volume as the cache size. 

\begin{figure}[!t] 
\centering
\begin{tabular}{@{\ }ccc}
\includegraphics[width=2in]{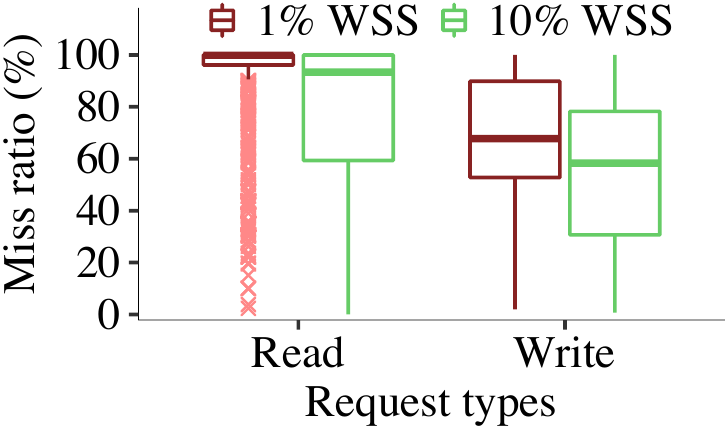} &
\includegraphics[width=2in]{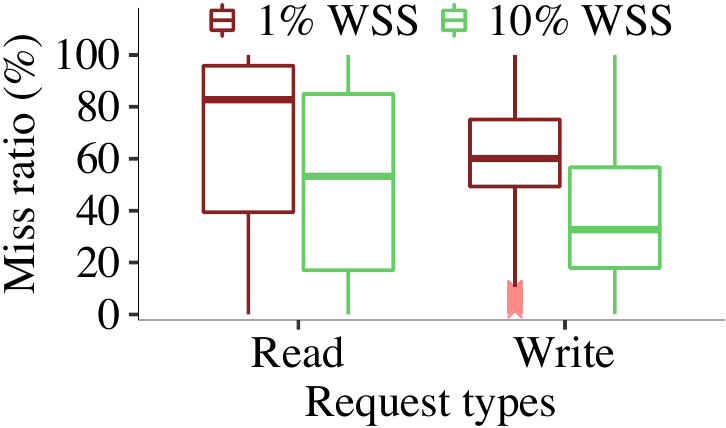} &
\includegraphics[width=2in]{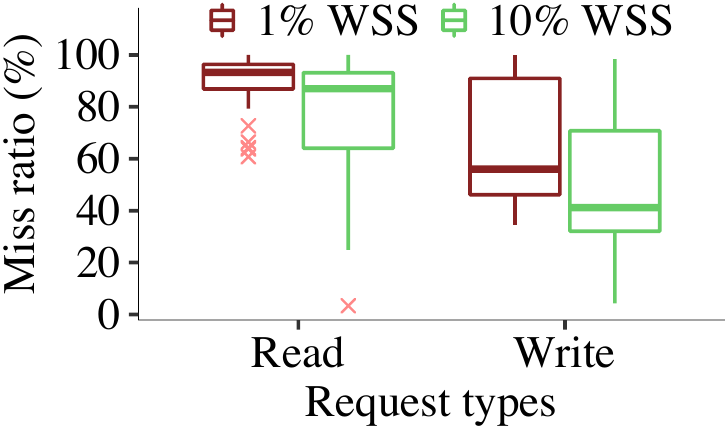} \\
{\small (a) AliCloud}  & 
{\small (b) TencentCloud}  & 
{\small (c) MSRC}
\end{tabular}
\caption{Finding~B.16: Boxplots of miss ratios for reads and writes across all
volumes, under the cache sizes of 1\% and 10\% of the WSS of a volume.} 
\label{fig:miss_ratio}
\end{figure}

Figure~\ref{fig:miss_ratio} shows the boxplots of miss ratios across all
volumes in all three traces. Some volumes show low miss ratios
(i.e., LRU-based caching is effective). For the cache size of 10\% of WSS, the
25th percentiles of the miss ratios for reads and writes are 59.4\% and 30.7\%
in AliCloud, respectively (Figure~\ref{fig:miss_ratio}(a)), while the
corresponding miss ratios are 17.0\% and 17.9\% in TencentCloud, respectively
(Figure~\ref{fig:miss_ratio}(b)), and 64.1\% and 32.0\% in MSRC, 
respectively (Figure~\ref{fig:miss_ratio}(c)). Also, some volumes in
TencentCloud can have very low miss ratios when the cache size is only 1\% of
WSS, implying that the access patterns of such volumes have high temporal
locality, while AliCloud and MSRC have fewer such volumes.  The low miss ratios
in TencentCloud suggest that we can potentially improve the read and write
performance using a small-size cache.  Such findings are also consistent with
those in \cite{zhang20osca}, which shows low miss ratios of some selected
volumes with small-size cache.

AliCloud has the highest absolute reductions in read miss
ratios when the cache size increases from 1\% to 10\% of WSS among all three
traces; for write miss ratios, TencentCloud shows the highest absolute
reductions for increased cache sizes. In AliCloud, the 25th percentiles of the
miss ratios for reads and writes reduce from 96.1\% to 59.4\% and from 52.8\%
to 30.7\% (i.e., 36.7\% and 22.1\% of absolute reduction), respectively
(Figure~\ref{fig:miss_ratio}(a)).  In TencentCloud, the 25th
percentiles of the miss ratios for reads and writes reduce from 39.4\% to
17.0\% and from 49.3\% to 17.9\% (i.e., 22.4\% and 31.4\% of absolute
reduction, Figure~\ref{fig:miss_ratio}(b)), while in MSRC, the 25th
percentiles of the miss ratios for reads and writes reduce from 86.9\% to
64.1\% and from 46.2\% to 32.1\% (i.e., 22.8\% and 14.1\% of absolute
reduction), respectively (Figure~\ref{fig:miss_ratio}(c)). 
In short, in all three traces, AliCloud and TencentCloud have the highest
reductions in read and write miss ratios, respectively, implying the 
significance of enlarging the cache size in cloud block storage workloads.

\subsection{Similarities and Differences Between Two Cloud Block Storage Traces}
\label{subsec:diff}

We highlight the major similarities and differences between the
two cloud block storage traces, AliCloud and TencentCloud. 

\paragraph{Load intensities. } AliCloud and TencentCloud show similar
intensities of volumes (Finding~B.1), but different observations
in burstiness and activeness.
\begin{itemize}[leftmargin=*]
\item 
In terms of burstiness, TencentCloud has lower overall burstiness than
AliCloud, and also has a lower fraction of volumes with high burstiness ratios
(Finding~B.2).  Nevertheless, both AliCloud and TencentCloud have more diverse
burstiness across volumes than MSRC (Finding~B.3).  
\item 
In terms of activeness, TencentCloud has higher activeness than AliCloud, in
both the number of active volumes and the active time in each volume
(Finding~B.5).  While writes are the dominant factor in the activeness of both
cloud block storage traces (Finding~B.6), TencentCloud is more read-active than
AliCloud (Finding~B.7). 
\item 
In terms of the distribution of traffic per day, unlike MSRC, the traffic of
both AliCloud and TencentCloud is almost evenly spread across daytime and
nighttime.  While AliCloud shows substantial read traffic at midnight,
TencentCloud does not (Finding~B.8).
\end{itemize}

\paragraph{Spatial patterns. } While AliCloud and TencentCloud show higher
randomness than MSRC, TencentCloud generally has higher randomness in I/O
requests and higher levels of traffic aggregation in blocks than AliCloud. 
\begin{itemize}[leftmargin=*]
\item 
In terms of the randomness in I/Os, the volumes in both AliCloud and
TencentCloud have higher percentages of random I/Os than those in MSRC.
Compared with AliCloud, TencentCloud generally shows a higher percentage of
random I/Os. For example, TencentCloud has a higher fraction of random
requests in half of the volumes than AliCloud (42.1\% versus 33.5\%)
(Finding~B.9). 
\item 
In terms of the aggregation of traffic, TencentCloud shows more traffic in
top-1\% and top-10\% of the most access-intensive blocks than AliCloud,
indicating that TencentCloud has a higher degree of traffic aggregation in a
small set of blocks (Finding~B.10).  Also, while both AliCloud and
TencentCloud have higher percentages of writes in write-mostly blocks than
MSRC, TencentCloud has higher read and write traffic aggregations 
than AliCloud in read-mostly and write-mostly blocks, respectively
(Finding~B.11).  Furthermore, while both AliCloud and TencentCloud have higher
fractions of update-intensive volumes (i.e., the volumes with high percentages
of update traffic) than MSRC, TencentCloud has a higher percentage of
update-intensive volumes than AliCloud (Finding~B.12).
\end{itemize}

\paragraph{Temporal patterns. }  TencentCloud generally has lower access
intervals on blocks in reads, writes, and updates.  It also has lower miss
ratios than AliCloud under the same cache configurations.  
\begin{itemize}[leftmargin=*]
\item 
In terms of the time intervals in accessing the same blocks, both AliCloud and
TencentCloud generally have smaller WAW times than RAW times and smaller RAR
times than WAR times.  However, TencentCloud generally has lower RAW, WAW,
RAR, and WAR times than AliCloud (Findings~B.13 and B.14).  It indicates that
TencentCloud has more access-intensive workloads than AliCloud. 
\item 
In terms of the update intervals, TencentCloud has 90\% of its update intervals
smaller than 2.0~hours, and its update intervals are generally smaller than
AliCloud (Finding~B.15).  
\item 
In terms of miss ratios, TencentCloud has lower miss ratios under the same
cache configurations. It has lower absolute reductions in read miss ratios than
AliCloud when the cache size increases from 1\% to 10\% of WSS. On the
contrary, it has higher absolute reductions in write miss ratios than AliCloud
when the cache size increases from 1\% to 10\% of WSS (Finding~B.16).
\end{itemize}

\subsection{Summary of Findings}
\label{subsec:summary} 

Finally, we discuss the implications of our findings of the trace analysis in
AliCloud, TencentCloud, and MSRC.  We show how the findings address
the design considerations for cloud block storage, including load balancing,
cache efficiency, and storage cluster management
(\S\ref{subsec:motivation}).

\paragraph{Load balancing.} We focus on the average and peak intensities as
well as the activeness of volumes.  From Finding~B.1, we observe that while
many applications are hosted in the cloud, the volumes in cloud block storage
(i.e., AliCloud and TencentCloud) have similar average load intensities to
those in traditional data centers (i.e., MSRC) which were monitored more than a
decade ago; however, the peak intensities are generally smaller. 

From Findings~B.2-B.4, we observe the existence of burstiness in a
non-negligible fraction of volumes.  While the overall burstiness remains low,
the burstiness can be severe in individual volumes across many types of
applications \cite{riska06}, indicating that these volumes may be provisioned
for high peak intensities but most of the bandwidth resources remain unused
\cite{narayanan08}. The high burstiness may hence lead to performance
degradations if load balancing is not properly maintained.
Furthermore, the higher diversity of workload burstiness makes load balancing
in cloud block storage more challenging than in traditional data centers.
Failing to deal with load imbalance and the diversity of workloads may cause
problems to the physical devices in cloud block storage, such as higher flash
failure rates \cite{xu19}.  Applying shared logs or distributed caches can
ease the load imbalance among volumes \cite{liu19distcache, xu19}.  Also,
the burstiness, as shown by the short inter-arrival times, suggests that I/O
requests tend to arrive in groups and can be further exploited to improve I/O
performance \cite{hsu03}. 

From Findings~B.5-B.7, we observe that writes are the dominant factor of
activeness in all three traces.
In particular, most volumes in cloud block storage (i.e., AliCloud and
TencentCloud) are write-dominant (\S\ref{subsec:highlevel}).  The differences
of activeness in reads and writes indicate that removing writes can produce a
high level of idle periods, so it is possible to offload writes (e.g., by
redirecting writes to other storage locations) to create idle periods in cloud
block storage workloads for power savings \cite{narayanan08}.

From Finding~B.8, I/O traffic is evenly spread across both daytime and
nighttime in AliCloud and TencentCloud.  This challenges background task
scheduling (e.g., garbage collection, defragmentation, and flushing caches
\cite{hsu03, riska06}) in cloud block storage. For example, performing
background tasks only at nighttime may be ineffective to reduce the
interference with foreground I/Os.  A careful design of I/O
scheduling for background and foreground activities becomes necessary for cloud
block storage.  In particular, we observe large read spikes near
midnight on a daily basis in some of the volumes in AliCloud.  How to prevent
such I/O spikes from interfering with cloud block storage system as a whole
needs careful attention.

Concerning the design of load balancing, the data placement
strategies should be aware of the diversity of workloads, the burstiness of
individual volumes, and the traffic distribution over time.  The log-structured
design \cite{rosenblum92} is proven useful for balancing the write traffic in
cloud-scale flash-based storage \cite{xu19}.

\paragraph{Cache efficiency.}  We study the spatial and temporal
characteristics of volumes, which provide guidelines for motivating new
caching designs for cloud block storage. 

From Findings~B.10 and B.16, we observe the patterns of both spatial and
temporal traffic aggregations in a small fraction of blocks, especially for
writes.  TencentCloud shows a stronger traffic aggregation compared with
AliCloud and MSRC.  Many volumes in cloud block storage show high aggregations
of reads and writes, implying that it is viable to allocate limited cache
resources for absorbing substantial amounts of reads and writes.

From Finding~B.11, we observe that many volumes in cloud block storage have
reads and writes aggregated in read-mostly and write-mostly blocks,
respectively.
Thus, one possible caching admission policy is to identify the read-mostly and
write-mostly blocks in workloads, as such blocks can absorb a substantial
amount of I/O traffic.  Also, the read-mostly and write-mostly blocks can be
put into different devices to separately reduce the read and write latencies
\cite{li16}.

From Findings~B.13 and B.14, the blocks that have been written tend to be
rewritten again.
In contrast, the blocks that have been read tend to receive another write after
a long period of time.  Thus, if our goal is to absorb writes with caching, a
possible strategy is to favor the caching of the blocks that have been written
rather than those that have been read, as the latter may unlikely generate
write hits.  Also, cloud block storage can benefit from disk-based write
caching \cite{soundararajan10}, due to the limited reads from the disk-based
cache.

\paragraph{Storage cluster management.}  Characterizing the spatial and
temporal characteristics of volumes is also critical for storage cluster
management.  Here, we focus on flash-based storage
(\S\ref{subsec:arch}). 

From Finding~B.9, we observe that upper-layer applications in cloud block
storage issue a high fraction of small and random I/Os, which are known to
hurt both the performance and endurance of flash-based storage \cite{min12}.
The log-structured storage design \cite{rosenblum92} and I/O clustering
\cite{min12} can help mitigate the overhead of small and random I/Os.

From Findings~B.12 and B.15, updates are common and have high variations
across volumes, both spatially and temporally.  The varying update coverage
across different volumes requires the underlying caches to use adaptive
caching methods to absorb update traffic. 
Also, the varying update patterns can harm the effectiveness of garbage
collection and wear leveling in flash \cite{he17}. Thus, cloud block storage
systems should take into account the varying patterns when optimizing
update workloads for flash-based storage.  A possible direction is to maintain
the flash-translation layer (FTL) at the system level \cite{chiueh14} to
flexibly coordinate the I/Os issued to flash.

From Finding~B.13, a larger number of WAW requests than RAW requests in
AliCloud and TencentCloud indicates that the next issued requests to newly
written blocks tend to be writes instead of reads.
If these written blocks are replicated across different nodes, we may choose
to update only one copy and invalidate other copies, instead of updating all
copies, in order to save the update overhead since the written data is likely
to be rewritten again \cite{hsu03}. 

\paragraph{Unexpected results.}  Finally, we highlight some unexpected results
reported in our findings.  

From Finding~B.6, we observe that the activeness of MSRC is dominated by
writes, yet the MSRC is read-dominant (\S\ref{subsec:highlevel}).  The results
indicate that MSRC is write-dominant from the perspective of activeness, but
is read-dominant from the perspective of the amount of I/O traffic. 

From Finding~B.8, we observe read spikes near midnight in both AliCloud and
MSRC traces.  In the corresponding read requests of the spike period, the
average read request size in AliCloud is larger than 360\,KiB, while that in
MSRC is smaller than 43.1\,KiB.  We suspect that there exist scheduled scan
activities in the corresponding volumes of AliCloud. 

From Finding~B.11, the limited aggregation of writes in write-mostly blocks in
MSRC is inconsistent with prior work \cite{li16}, which emphasizes that most
of the write requests access write-mostly blocks. The reason is that the
previous study \cite{li16} considers 12 volumes, while we consider all 36
volumes instead. 

\section{Related Work}
\label{sec:related}

We review related work on the field studies on storage workloads and how they
inspire storage system designs. 

\paragraph{Characterization of storage workloads.} Several field studies
characterize storage workloads using block-level I/O traces in various
architectures, such as consumer electronics \cite{riska06},
virtual machines \cite{ahmad07}, Windows servers \cite{kavalanekar08,
narayanan08}, smartphone applications \cite{zhou15}, containerized applications
\cite{harter16}, and virtual desktop infrastructures \cite{lee17}.
Yadgar et al. \cite{yadgar21} perform I/O workload analysis and study the
performance implications (e.g., read/write amplifications and flash read
costs) for SSD-based storage. In contrast, our field study focuses on cloud
block storage that supports a diverse set of cloud applications in large-scale
production. In particular, we provide findings and insights on performance
optimizations for load balancing, caching efficiency, and storage cluster
management.

Table~\ref{tab:compare} summarizes the traces used in the existing block-level
trace studies \cite{ahmad07, kavalanekar08, narayanan08, zhou15, harter16,
lee17, zhang20osca, yadgar21} in the literature, in terms of the number of
traces (or volumes in our case), the trace durations, the number of read and
write requests, and the total data traffic of reads and writes.  Our trace
analysis has the largest scale compared with the existing block-level trace
studies at the time of the writing.  Ahmad et al.  \cite{ahmad07} do not show
the overall statistics but mention that the total I/O size is at most 10\,GiB.
Harter et al. \cite{harter16} only focus on reads, and their analysis comprises
of 57 docker images, each of which has the average read traffic of 27\,MiB.
Zhang et al. \cite{zhang20osca} collected the TencentCloud traces, but they
mainly focus on the cache allocation design based on trace-driven evaluation
instead of providing detailed trace analysis.

\begin{table}[t]
\small
\centering
\renewcommand{\arraystretch}{1.1}
\setlength\tabcolsep{6pt}
\begin{tabular}{c|c|c|c|c|c|c|c}
\hline
& {\bf \begin{tabular}{@{}c@{}} MSRC \\ \cite{narayanan08} \end{tabular} } & 
{\bf \begin{tabular}{@{}c@{}} MS-Prod\\ \cite{kavalanekar08} \end{tabular} } & 
{\bf \begin{tabular}{@{}c@{}} Zhou et \\ al. \cite{zhou15} \end{tabular} } & 
{\bf \begin{tabular}{@{}c@{}} Lee et \\ al. \cite{lee17} \end{tabular} } & 
{\bf \begin{tabular}{@{}c@{}} Tencent \\ -Cloud \cite{zhang20osca} \end{tabular} } & 
{\bf \begin{tabular}{@{}c@{}} SSDTrace \\ \cite{yadgar21} \end{tabular} } & 
{\bf AliCloud} \\ 
\hline
\hline
{\bf \#Volumes}               & 36    & 43   & 25 & 321   & 4,995  & 1 & 1,000 \\ 
\hline
{\bf Duration (days)}         & 7     & 0.003-1  & < 0.34 & 28 & 9.04 & 0.40 & 31 \\ 
\hline
{\bf \#Reads (millions)}      & 304.9 & 126.2 & 0.04 & 2455.4 & 10,030.2 & 342.9 & 5,058.6 \\ 
\hline
{\bf \#Writes (millions)}     & 128.9 & 87.3  & 0.13 & 898.3  & 23,592.0 & 9.69 & 15,174.4 \\ 
\hline
{\bf Read Traffic (TiB)}      & 9.04  & 2.98  & 0.002 & 64.8 & 282.3 & 2.94  & 161.6 \\ 
\hline
{\bf Write Traffic (TiB)}     & 2.39  & 1.70  & 0.006 & 15.0 & 837.2 & 6.38  & 455.5 \\ 
\hline
\end{tabular}
\vspace{3pt}
\caption{Statistics of existing block-level trace studies 
\cite{kavalanekar08, narayanan08, zhou15, lee17, zhang20osca, yadgar21}.}
\label{tab:compare}
\end{table}

\paragraph{Inspirations from load intensity.}  Some designs are inspired by
the characteristics of load intensity in storage workloads.  Narayanan et al.
\cite{narayanan08} offload writes to reduce power consumptions with the
observation that some volumes are idle in reads, thereby removing writes in
those volumes can increase the idle periods for power saving.  SRCMap
\cite{verma10} reduces power consumptions using sampling and replication,
based on the observation on the I/O size and intensity of active data
sets.  Ursa \cite{li19} adopts the log-structured design, based on the
observation that small writes dominate in real-world workloads. 

\paragraph{Inspirations from spatial patterns.} Some designs exploit the
spatial characteristics of storage workloads.  BORG \cite{bhadkamkar09}
organizes frequently written data in a small dedicated disk partition to
reduce the I/O seek time.  FlashTier \cite{saxena12} manages sparse address
mappings in flash caching, as storage I/Os are often aggregated in a small
number of blocks. Desnoyers \cite{desnoyers12} proposes an analytical model
for cleaning algorithms in flash devices and analyzes the aggregation of
written blocks in specific working sets.
ACGR \cite{li16} regulates I/O accesses for flash storage, based on the
observation of read and write aggregations in read-only and write-only
blocks, respectively. To improve the update performance in erasure-coded
storage, CodFS \cite{chan14} proposes dynamic reserved space management for
parity updates to address the varying working sets of updates across storage
workloads, while PBS \cite{zhang20pbs} exploits the large fractions of
overwrites to mitigate parity update overhead.

\paragraph{Inspirations from temporal patterns.}  Some designs exploit the
temporal characteristics of storage workloads.  Griffin \cite{soundararajan10}
leverages the large time intervals between writes and the subsequent reads to
the same block to build an HDD-based write cache for improving the SSD
lifetime.  Arteaga et al. \cite{arteaga14} propose a cache-optimized RAID
technique to minimize the RAID overhead in cloud storage, based on the
comparisons between write-back caching and write-through caching on a set of
block I/O traces from production servers in the cloud. CloudCache
\cite{arteaga16} chooses the window size of the model based on the hit ratio
analysis on two-week traces in the cloud.  Some studies leverage the
characteristics of update intervals in storage workloads for improving write
performance \cite{liu12}, lifetime \cite{cai15}, garbage collection modeling,
and data reduction \cite{yang19} in SSDs. Counter Stacks \cite{wires14},
SHARDS \cite{waldspurger15}, and OSCA \cite{zhang20osca} consider the reuse
distance (i.e., the number of distinct items accessed between two accesses to
the same item) to improve caching efficiency. 

\paragraph{Cloud block storage systems.} Several cloud block storage designs
are proposed in the literature. Parallax \cite{meyer08} provides storage
virtualization for virtual machines atop shared block storage.  Blizzard
\cite{mickens14} manages POSIX applications atop cloud block storage. 
Ursa \cite{li19} is a hybrid block storage system that combines HDDs 
and SSDs for cloud-scale virtual disks.  PBS \cite{zhang20pbs} supports
erasure-coded cloud block storage with efficient updates.  Our recent work
SepBIT \cite{wang22} is a data placement scheme that mitigates the write
amplification of garbage collection in log-structured cloud block storage, and
its design and evaluation are based on the AliCloud and TencentCloud traces. 
In this work, we conduct an in-depth trace analysis that provides suggestions
for improving the cloud block storage design. 

\section{Conclusion}
\label{sec:conclu}

We present an in-depth comparative trace analysis on the production
block-level I/O traces at Alibaba Cloud (AliCloud), Tencent Cloud Block
Storage (TencentCloud), and Microsoft Research Cambridge (MSRC); the AliCloud
and TencentCloud traces are from cloud block storage systems, while the MSRC
trace is collected from enterprise data centers.  We reveal the commonalities
and differences of the three sources of traces.  We first identify 6
findings through the high-level analysis on the basic I/O statistics. We
also identify 16 findings through the detailed analysis, based on which we
further discuss the implications on three practical design considerations for
cloud block storage, including load balancing, cache efficiency, and storage
cluster management.

\bibliographystyle{abbrv}
\bibliography{reference}

\end{document}